\documentclass[preprint,aps]{revtex4}
\usepackage{graphicx}
\usepackage{dcolumn}
\usepackage{bm}
\usepackage{epsf}

\newcommand{\sbf}[1]{\mbox{{\scriptsize$\bf{#1}$}}}
\def\ast{{\dagger}}
\def\bx{{\bf x}}

\def\sbq{{\sbf q}}
\def\bq{{\bf q}}

\def\cE{{\cal E}}

\newcommand{\nwl}{\\[5mm]}
\newcommand{\edc}{\end{document}}
\newcommand{\bb} {\begin{thebibliography}}
\newcommand{\eb}{\end{thebibliography}}
\newcommand{\bi}[1]{\bibitem{#1}}
\newcommand{\bc}{\begin{center}}
\newcommand{\ec}{\end{center}}

\newcommand{\be}{\begin{equation}\small}
\newcommand{\ee}{\end{equation}\normalsize}
\newcommand{\bea}{\begin{eqnarray}}
\newcommand{\eea}{\end{eqnarray}}
\newcommand{\ba}{\begin{array}{l}   }
\newcommand{\lab}[1]{\label{#1}}
\newcommand{\ea}{\end{array}}
\newcommand{\Tr}{\mbox{Tr}}
\newcommand{\dsfrac}{\displaystyle\frac}
\newcommand{\ds} {\displaystyle}
\newcommand{\summa}{\ds\sum}

\newcommand{\re}[1]{(\ref{#1})}

\newcommand{\ci}{\cite}

\newcommand{\dsint}{\ds\int}

\def\bfk{{\bf k}}
\def\bfq{{\bf q}}
\def\bfx{{\bf x}}

\newcommand{{\vergul}}{  ,}

\newcommand{\veps}{\varepsilon }

\newcommand{\calv}{{{\cal V}}}

\newcommand{\uj}{{U/J}}
\newcommand{\tilmu}{\tilde{\mu}}

\begin{document}
\draft
\title{Quantum phase transitions in optical lattices beyond
Bogoliubov approximation}
\author{H. Kleinert$^{a}$}\email{h.k@fu-berlin.de}
\author{Z. Narzikulov$^{b}$}\email{narzikulov@inp.uz}
\author{Abdulla Rakhimov$^{a,b}$}\email{rakhimovabd@yandex.ru}
\affiliation{
$^a$ Institut f$\ddot{u}$r Theoretische Physik, Freie Universit$\ddot{a}$t Berlin, Arnimallee 14, D-14195 Berlin, Germany\\
$^b$Institute of Nuclear Physics, Tashkent 100214, Uzbekistan
}
\begin{abstract}

We study
quantum phase transition
from the superfluid to a  Mott insulator
 in optical lattices
using a Bose-Hubbard
Hamiltonian. For this purpose we have developed a field theoretical approach in
 terms of path integral formalism  to  calculate the second-order quantum corrections to the energy density as well as to the superfluid fraction in cubic optical lattices.
Using present approach the condensate fraction and ground state energy
are calculated as functions of the $s$-wave scattering length.
In contrast to the Bogoliubov
model, which is technically speaking a one-loop approximation,
we carry the
calculation up
to two loops, and improve the
result further by
variational perturbation theory.
The result suggests
that the quantum phase transition exists.


\end{abstract}
\maketitle
\section{Introduction}
 Optical lattices are known  as the
gases of ultracold atoms trapped
 in periodic potentials created by standing waves of laser light. The actuality of experimental
 and theoretical investigations of these artificial crystals
 bound by light can be justified by following two factors \ci{Morchrev}:

1) Neutral atoms in these  optical lattices have a number of
affective futures that make them interesting candidates for the
realization of a quantum computer \ci{rous2003}.

2) They may be used to stimulate various lattice models of fundamental
 importance to condensed matter physics to study in a controlled way
  in solid-state physics, since one is able to finely tune
   the properties and geometry of the lattices.
In particular, it is possible to control the Hamiltonian parameters
and study various regimes of interest. Similarly to the ordinary
Bose - Einstein Condensation (BEC) of gases, the quantum phase
transitions in optical lattices were first predicted theoretically
\ci{fisher}  and have
recently been observed experimentally \ci{Greiner}.

Most of the theoretical investigations are based on Bose-Hubbard Hamiltonian:
\begin{equation}
H=-J\sum _{<{\sbf i},{\sbf j}>}\hat{c}_{\sbf{i}}^{\dag} \hat{c}_{\sbf{j}}  +\frac{U}{2} \sum
_{\sbf{i}}^{N_s}\hat{c}_{\sbf{i}}^{\dag} \hat{c}_{\sbf{i}}^{\dag} \hat{c}_{\sbf{i}} \hat{c}_{\sbf{i}}
+\sum _{\sbf{i}}^{N_s}(\varepsilon _{\sbf{i}} -\mu )\hat{c}_{\sbf{i}}^{\dag} \hat{c}_{\sbf{i}}
\label{H11}
\end{equation}
where $\hat{c_{\sbf{i}}}^{\dag}$ and $\hat{c_{\sbf{i}}}$ are the bosonic creation
 and annihilation operators on the site $i$; the  sum over $<{\sbf i},{\sbf j}>$ includes only pairs of nearest neighbors;
  $J$ is the hopping amplitude, which is responsible for the
   tunneling of an atom from one site to another neighboring site;
    $U$ is the on site repulsion energy; $N_s$ - number of sites.
Presently it is well established that at very low temperature
 $(T\rightarrow 0)$   a system of bosons described by the Hamiltonian
  \re{H11} could be on superfluid (SF) or
  in Mott insulator (MI) phase. Clearly there would be a quantum phase
   transition between these two phase depending on parameters
    $U$ and $J$. Particularly, when the hopping term is  dominated,
     ${U}/{J}\ll1$, the system prefers to be in the SF phase.
      On the other hand when the repulsion prevails the kinetic term,
${U}/{J}\gg1$, the system would be in MI phase where
     each atoms is absolutely localized near a site.

Clearly the superfluid phase may consist not only of condensed
particles with a number $N_{0}$, but also of $N_{1}$  uncondensed ones,
whose sum
$N_0+N_1=N$  is the total number of particles.
 The critical interaction  strength
$\kappa_{\rm crit}\equiv{(U/J)}_{\rm crit}=29.34$ and $\kappa_{\rm crit}=3.6$, for $D=3$ and $D=1$
  respectively, of the quantum phase
 $SF \rightarrow MI$
 transition estimated by Monte Carlo calculations \ci{svistun,batrouni}
  at filling factor $\nu=1$
 is in good agreement with the experimental data.

To make easier further reading we clarify some specific features of
these two phases:\\
 SF phases is characterized by long-range
correlation, a continuous (gapless) excitation spectrum and a finite
compressibility. Since there exists a condensate with a number of particles
$N_{0}\neq0$, the gauge symmetry is spontaneously broken in
accordance with Bogoliubov and Ginibre theorems.
In contrast, in the Mott insulator phase, there is no long-range
correlation neither breaking of gauge symmetry. The excitation
spectrum has a gap and the system is incompressible, since there is
a fixed  number of atoms per-site. This new state of
matter can survive only at zero temperature and integer filling
factor $\nu$.

It is interesting to note that there are two kinds of experiments
observing above quantum phase transition, depending on the starting
point. In the experiments by Greiner et al \ci{Greiner} one first
creates a BEC in a conventional harmonic trap and then adiabatically
adds the periodic optical potential. In the second method, pioneered
by the Florence group \ci{Burger} one uses a conventional protocol
for evaporative cooling in a magnetic trap down to temperatures just
above the threshold for BEC. At this point the optical lattice
potential is switched on and evaporative cooling continues. In this
way, the system condenses directly into a ground state of the
harmonic plus periodic potential. It seems to be that the first
method is good to observe SF$\rightarrow$MI while the second one is
good for MI$\rightarrow$SF transitions.

Similarly, most of theoretical approaches can be divided into two
classes: SF$\rightarrow$MI and MI$\rightarrow$SF ones. The latter
are based on the Ginzburg - Landau theory as describes
for instance in Ref.~\ci{Pelster}.
They are  well suited to analyze the time-of light
pictures and the resulting visibility at zero and finite temperatures.
In the former class (SF$\rightarrow$MI) one uses
a perturbative scheme  \ci{stoofbook}
within a decoupling (or single site)
approximation  due to Gutzwiller. This variational appoach which was first proposed for a fermion system
 \ci{Gutzwiller}, and further developed for bosons in Refs.~\ci{RJ,Krauth},
has the
following drawbacks  \ci{Yukalovobsor}
(see also last lines of  Sec. IV):
\begin{itemize}
\item
 The mean field Hamiltonian which
    features single boson terms does not conserve the total number of bosons
    \ci{vezzani};
    \item
    Tunneling of uncondensed atoms is neglected;
    \item
The critical value $\kappa_{\rm crit}$ does not depend on the lattice dimension.
\end{itemize}
    Nevertheless, the prediction of decoupling  approximation for
      $\kappa_{\rm crit}=34.98$ at filling factor $\nu=1$ is in agreement with
the well
      established value given above.
          Some years ago
an application of the Hartree-Fock-Popov  approximation (which is
widely used to study BEC of atomic gases and even triplons
\ci{ourANNALS}) to
optical lattices was presented by Stoof et al. \ci{Stoofmakola}.
Studying the dependence of the condensate number $N_{0}$ on $\kappa=U/J$, i.e.
$N_{0}(U/J)$ they observed that $N_{0}$ never reaches zero for
finite values of $\kappa$, implying that this
approximation is unable to predict a possible phase transition to a
Mott-insulator phase.   Moreover, a
Hartree-Fock-Bogoliubov (HFB) approximation applied to the
Bose-Hubbard Hamiltonian gives no quantum phase transition for optical lattices \ci{ourYUK}. Hence we find it
interesting to study the possibility of such a transition
if we go beyond these approximations.

 In the present work   we shall investigate BEC in optical lattices by applying
a two-loop approximation and treating the result
by
    variational perturbation theory (VPT) \ci{KLVPT}.
       It will be shown that,
while the ground
       state energy is rather sensitive to the filling
       factor in commensurate situations,
this is not so for
 arbitrary
condensate fractions $ n_0=N_0/N$.
We find that
 $ n_0$ goes to zero at $\kappa\sim 6 \div 6.5 $
for $\nu=1,2,3$ in  $D=3$ dimensions.
In  $D=1$ dimension, this happens at
$\kappa\sim 4 $.

The plan of this paper is as follows.
 In Sec. II  the basic equations in functional formalism
 for Bose-Hubbard Hamiltonian are formulated.
  In Sec. III we derive explicit expressions
   for the effective potential in two-loop order. In Sec. IV we obtain
    condensate fraction
   vs input parameters $U,J, \nu$ .
    The quantum corrections to the
   energy of the system is discussed in Sec. V.
   In  Sec. VI we present numerical results and discussions.
  The last Sec. VII summaries our results.
\section{The action and propagators in Bose-Hubbard model}
 The action at zero temperature, ($T=0$) that describes a gas of atoms in a periodic potential is given by
 \begin{eqnarray}
  S(\varphi^{\ast},\varphi)&=&\int dt d\bfx \Bigg[\varphi^{\ast}i\partial_{t}\varphi+\varphi^{\ast}
   \frac{\vec{\nabla}^2}{2m}\varphi+\mu\varphi^{\ast}\varphi-V_{\rm ext}({\bfx})
   \varphi^{\ast}\varphi\Bigg]\nonumber\\
    &&{}-\frac{1}{2}\int\varphi^{\ast}(\bfx)\varphi^{\ast}(\bfx^{\prime})V(\bfx-\bfx^{\prime})
    \varphi(\bfx)
    \varphi(\bfx^{\prime})dt
     d{\bf x} d{\bf x}^{\prime}
     \label{1.1}
      \end{eqnarray}
where the isotropic optical lattice potential is described by \ci{Greiner}
\begin{eqnarray}
 V_{\rm ext}({\bf x})=V_{0}\sum_{\alpha=1}^{D}\sin^{2}
 \left(
 \frac{2\pi{ x_{\alpha}}}{\lambda}
 \right)\label{1.2}
  \end{eqnarray}
with $\lambda$ the wave length of the laser light.
The lattice points
lie at  the positions \cite{GFCM}
\begin{eqnarray}
{\bf x}_{\sbf i}={\bf i}\,a,
\label{@}\end{eqnarray}
where $a$ is the lattice spacing, and
\begin{eqnarray}
{\bf i}\equiv
(i_1,i_2,\dots,i_d)
\label{@}\end{eqnarray}
are integer-valued vectors.
 It can be shown \cite{stoofbook,Yukalovobsor}
 that
the Wannier representation of the
 Hamiltonian corresponding to the action (\ref{1.1}) is equivalent to well known Bose-Hubbard model          \re{H11}.

The on-site energy, $\varepsilon_{\sbf{i}}$, the amplitude of hopping -- $J$ and on-site interaction strength $U$ are related to $V_{\rm ext}({\bf{x}})$  and $V
({\bf{x}}-{\bf{x}}^{\prime})$ as follows:
\begin{eqnarray}
 \varepsilon_{\sbf{i}}&=&\int d{\bf{x}}\omega^{\ast}_{0}({\bf{x}}-{\bf{x}}_{\sbf{i}})
  \left\{-\frac{\hbar^{2}\nabla^{2}}{2m}+V_{\rm ext}({\bf{x}})\right\}
   \omega_{0}({\bf{x}}-{\bf{x}}_{\sbf{i}})\label{1.4}\\
 J_{{\sbf i},{\sbf j}}&=&-\int d {\bf{x}}\omega^{\ast}_{0}({\bf{x}}-{\bf{x}}_{\sbf{i}})
  \left\{-\frac{\hbar^{2}\nabla^{2}}{2m}+V_{\rm ext}({\bf{x}})\right\}
   \omega_{0}({\bf{x}}-{\bf{x}}_{\sbf j})\label{1.41}\\
 U&=&\int d{\bf{x}}\int d{\bf{x}}^{\prime}\omega^{\ast}_{0}({\bf{x}}-
  {\bf{x}}_{\sbf{i}})\omega^{\ast}_{0}({\bf{x}}-{\bf{x}}_{\sbf{i}})V({\bf{x}}-{\bf{x}}')
   \omega_{0}({\bf{x}}^{\prime}-{\bf{x}}_{\sbf{i}})\omega_{0}({\bf{x}}^{\prime}
    -{\bf{x}}_{\sbf{i}})\label{1.5}
    \end{eqnarray}
where $\omega_{n}({\bf x})$ are Wannier functions.
In the tight-binding limit and pseudopotential approximation, $
V({\bf{x}}-{\bf{x}'})=4\pi a \delta({\bf{x}}-{\bf x}')/m$
the equations (\ref{1.41}), (\ref{1.5}) are simplified as:
\begin{eqnarray}
 J&=&\frac{4}{\sqrt{\pi}}E_{r}\left(\frac{V_{0}}{E_{r}}\right)^{3/4}
  \exp\left\{-2\left(\frac{V_{0}}{E_{r}}\right)^{1/2}\right\}\label{1.6}\\
 U&=&\frac{2\pi\omega a}{l \sqrt{2\pi}}\label{1.61}
  \end{eqnarray}
  where $E_{r}=2\pi^{2}/m\lambda^{2}$, $a$ is the s-wave scattering length, and   $l=\sqrt{1/m\omega}=(E_{r}/V_{0})^{1/4}\lambda/4\pi$ is the harmonic oscillator length.

In terms of parameters $J$ and $U$ the action (\ref{1.1}) can be rewritten as follows:
\begin{eqnarray}
 S(\varphi^{\ast},\varphi)&=&\int dt\Bigg\{\sum_{\sbf{i}}\varphi^{\ast}({\bf x}_{\sbf i}, t)[i\partial_{t}+\mu]
  \varphi({\bf x}_{\sbf i}, t)+J\sum_{<{\sbf i},{\sbf j}>}\varphi^{\ast}
({\bf x}_{\sbf i}, t)\varphi({\bf x}_{\sbf j}, t)\nonumber\\
   &&{}-\frac{U}{2}\sum_{\sbf{i}}\varphi^{\ast}({\bf x}_{\sbf i}, t)\varphi^{\ast}({\bf x}_{\sbf i}, t)
    \varphi({\bf x}_{\sbf i}, t)\varphi({\bf x}_{\sbf i}, t)\Bigg\}\label{1.7}
     \end{eqnarray}
The grand-canonical partition function $Z$, and the effective  potential at zero temperature,  ${\cal{V}}$, can be found as \ci{jackiw}:
\begin{eqnarray}
 Z&=&\int{\cal{D}}\varphi^{\ast}{\cal{D}}\varphi \displaystyle{e^{iS(\varphi^{\ast},\varphi)}}\label{1.8}\\
  {\cal{V}}&=&\frac{i}{T}\ln{Z}\label{1.9}
   \end{eqnarray}
where $\int dt=T$ is the total time interval. Note that, in accordance with
the background field method  \ci{oursinggap}, which will be used below,
in evaluation of the effective potential only connected single - particle
irreducible Feynman diagrams should be included.
The ground state expectation   value of an operator $\hat{A}(\varphi^{\ast},\varphi)$
 can be expressed as a functional integral:
\begin{eqnarray}
 \langle \hat{A}\rangle=\frac{1}{Z}\int{\cal{D}}\varphi^{\ast}{\cal{D}}\varphi
  \hat{A}(\varphi^{\ast},\varphi)e^{iS(\varphi^{\ast},\varphi)}\label{1.10}
   \end{eqnarray}
At zero temperature the system could undergo into BEC state. The necessary and sufficient condition for Bose-Einstein condensation is the spontaneous gauge-symmetry breaking which is established by Bogoliubov shift \ci{Yukalovobsor}:
\begin{eqnarray}
 \varphi({\bf x}_{\sbf i},t)=\sqrt{\nu  n_0}+\tilde{\varphi}
({\bf x}_{\sbf i},t)\label{1.11}
  \end{eqnarray}
where $\nu=N/N_{s}$-- filling factor,and the condensate fraction, $ n_0=N_{0}/N$, is constant
for regular lattice without magnetic trap.

Substituting (\ref{1.11}) into (\ref{1.7}) and parameterizing  quantum field  $\tilde{\varphi}({\bf x}_{\sbf i}, t)$ in terms of two real-valued quantum fields $\varphi_{1}({\bf x}_{\sbf i}, t)$ and $\varphi_{2}({\bf x}_{\sbf i}, t)$ as
\begin{eqnarray}
 \tilde{\varphi}({\bf x}_{\sbf i}, t)&=&\frac{1}{\sqrt{2}}
  (\varphi_{1}({\bf x}_{\sbf i},t)+i\varphi_{2}({\bf x}_{\sbf i}, t))\nonumber\\
  \tilde{\varphi}^{\ast}({\bf x}_{\sbf i}, t)&=&\frac{1}
   {\sqrt{2}}(\varphi_{1}({\bf x}_{\sbf i},t)-i\varphi_{2}({\bf x}_{\sbf i}, t))\label{1.12}
   \end{eqnarray}
  one may separate the action as follows
\begin{eqnarray}
 S&=&S^{0}+S^{(1)}+S^{(2)}+S^{(3)}+S^{(4)}\label{1.13}\\
  S^{0}&=&N_{s}\int dt\left[\mu\nu  n_0+Jz_0\nu
    n_0-\frac{U}{2}\nu^{2}n_{0}^{2}\right] \label{1.14}\\
    S^{(1)}&=&\sqrt{2\nu  n_0}\bigg[Jz_{0}+\mu-U\nu  n_0\bigg]\int dt\sum_{\sbf{i}}\varphi_{1}
     ({\bf x}_{\sbf i}, t)\label{1.15}\\
      S^{(2)}&=&\frac{1}{2}\int dt\sum_{\sbf{i}}\sum_{a,b=1,2}\bigg[-\varepsilon_{ab}
       \varphi_{a}({\bf x}_{\sbf i}, t)\partial_t\varphi_{b}({\bf x}_{\sbf i}, t)-\varphi_{a}({\bf x}_{\sbf i}, t)X_{a}\varphi_{b}({\bf x}_{\sbf i}, t)
        \delta_{ab}\bigg]\nonumber\\
         &&{}+\frac{J}{2}\int dt\sum_{<{\sbf i},{\sbf j}>}\sum_{a=1,2}\varphi_{a}({\bf x}_{\sbf i}, t)\varphi_{a}
          ({\bf x}_{\sbf j}, t)\label{1.16}\\
     S^{(3)}&=&-\frac{U\sqrt{2\nu  n_0}}{2}\int dt\sum_{\sbf{i}}\bigg[\varphi_{1}({\bf x}_{\sbf i}, t)
      \varphi_{2}^{2}({\bf x}_{\sbf i}, t)+\varphi^{3}_{1}({\bf x}_{\sbf i}, t)\bigg]\label{1.17}\\
     S^{(4)} &=&-\frac{U}{8}\int dt\sum_{\sbf{i}}\bigg[\varphi_{1}^{4}({\bf x}_{\sbf i},t)+\varphi_{2}^{4}({\bf x}_{\sbf i},t)
      +2\varphi_{1}^{2}({\bf x}_{\sbf i},t)\varphi_{2}^{2}({\bf x}_{\sbf i},t)\bigg]
      \label{1.18}.
 \end{eqnarray}
 In (\ref{1.16}) $\varepsilon_{ab}$ is the antisymmetric tensor with $\varepsilon_{12}=1,\quad \varepsilon_{21}=-1$, and
 \begin{eqnarray}
  X_{1}&=&-\mu+3U\nu  n_0\nonumber\\
   X_{2}&=&-\mu+U\nu  n_0\label{1.181}
    \end{eqnarray}

For a homogenous system the condensate is uniform and it is convenient to decompose
the fluctuations into a Fourier series \ci{HK,danshita}
\begin{eqnarray}
 \varphi_{a}({\bf x}_{\sbf j}, t)=\dsfrac{1}{\sqrt{N_{s}^d}}\sum_{\sbf q}
 \int\dsfrac{d\omega}
  {(2\pi)}\varphi_{a}(\vec{q},\omega)e^{-i\omega t}\exp\left[\frac{2i\pi {\bf j}}
{N_s}{\bf q}\right]
  \label{1.19}
   \end{eqnarray}
 where ${\bf q}=\{q_1, q_2 \ldots q_d\}$ with $q_i$ running from $1$ to $N_s-1$
 is an integer-valued vector field
associated with all wave vectors in the Brioullin zone:
${\vec q}= 2\pi\,{\bf q}/a$, and
\be
\frac{1}{N_{s}}\sum_{{\sbf q}}\equiv
 \frac{1}{N_{s}^{d}}\sum_{q_1=1}^{N_s-1}\sum_{q_2=1}^{N_s-1}\ldots \sum_{q_d=1}^{N_s-1}.
\ee
 The $\vec{q}=0$ mode, i.e. the Goldstone mode,
is omitted from the sum,
to achieve orthogonality between the condensate
 and noncondensed modes.
 In momentum space  the quadratic term $S^{(2)}$ as follows:
\begin{eqnarray}
 S^{(2)}&=&\frac{1}{2}\int\sum_{{\sbf q},{\sbf q}^{\prime}}\varphi_{a}(\bq,\omega)
M_{ab}
  (\bq,\omega,\bq^{\prime},\omega^{\prime})\varphi_{b}(\bq^{\prime},\omega^{\prime})\frac{d\omega
   d\omega^{\prime}}{(2\pi)^{2}}
   \lab{s23}
   \\
 M_{11}(\bq,\omega,\bq^{\prime},\omega^{\prime})&=&-[X_{1}+{\varepsilon}(\bq)
  -Jz_0]\delta(\omega+\omega^{\prime})\delta_{\bfq,-\bfq^{\prime}},\quad M_{12}(\bq,\omega,\bq^{\prime},\omega^{\prime})=i\omega,
  \label{1.1911}
  \\
 M_{22}(\bq,\omega,\bq^{\prime},\omega^{\prime})&=&-[X_{2}+{\varepsilon}(\bq)
  -Jz_0]\delta(\omega+\omega^{\prime})\delta_{\bfq,-\bfq^{\prime}},\quad M_{21}(\bq,\omega,\bq^{\prime},\omega^{\prime})=-i\omega,
  \label{1.191}
  \end{eqnarray}
with $z_0$ being the number of nearest neighbors.
From this we extract
the Fourier transformation of the propagator of the
fields $\varphi_{1}$, and $\varphi_{2}$ as the $2\times2$ matrix:
  \begin{eqnarray}
   G(\omega,\bq)=\frac{i}{\omega^{2}-\cE^{2}(\bq)+i\epsilon}
    \left(
\begin{array}{lr}
 X_{2}+{\varepsilon}(\bq)-Jz_0 & -i\omega\\
 i\omega &X_{1}+{\varepsilon}(\bq)-Jz_0
\end{array}\right)\label{1.20}
  \end{eqnarray}
where
\begin{eqnarray}
 \cE(\bq)&=&\sqrt{(X_{1}+{\varepsilon}(\bq)-Jz_0)
  (X_{2}+{\varepsilon}(\bq)-Jz_0)}\nonumber\\
   {\varepsilon}(\bq)&=&2J\bigg(d-\summa_{\alpha=1}^{d}\cos (2\pi q_{\alpha}/N_s)\bigg)
   \label{1.21}
    \end{eqnarray}
In coordinate space for a regular lattice the propagator is translational
invariant
\be
\ba
 {G}_{ab}{({\bf x}_{\sbf i},t;{\bf x}_{\sbf j},t')}\equiv
   {G}_{ab}{({\bf x}_{\sbf i}-{\bf x}_{\sbf j},t-t')}=\langle\varphi_{a}({\bf x}_{\sbf i},t)
   \varphi_{b}({\bf x}_{\sbf j},t')\rangle\label{1.22}
\ea
 \ee

 Note that, in deriving  \re{s23}-\re{1.21}, the following relations
have been used:
 \be
 \ba
 \ds\sum_{<{\sbf m},{\sbf j}>}\exp\left[\frac{i2\pi}{N_s}({\bf j}\cdot{\bf q}-
{\bf m}\cdot{\bf p}
)\right]=
 2N_s\delta_{\bfq,{\bf p}}\ds\sum_{\alpha=1}^{d}
 \cos (2\pi q_\alpha/N_s), \\
 \ds\sum_{\sbf j}\exp\left[\frac{i2\pi {\bf  j}}{N_s}{(\bf q}-{\bf p})\right]=
 N_s\delta_{{\bfq},{\bf p}}\\
 \sum_{<{\sbf i},{\sbf j}>} [1]=z_0=2d, \quad \quad \sum_{{\bf q}} [1]=
N_s, \quad\quad \sum_{\sbf{i}} [1]=N_s.
  \ea
 \ee


\section{The effective potential in two-loop approximation}

To organize the quantum corrections in a two-loop expansion, we separate the terms in the action (\ref{1.13}) into a free part and  interaction parts following
Jackiws pioneering work \ci{jackiw}
\bea
 S&=&S_{\rm cl}+S_{\rm free}+S_{\rm int}\label{11.1}\\
  S_{\rm cl}&=& S^{0}=N_{s}\int dt\left\{\mu\nu  n_0+Jz_0\nu
    n_0-\frac{U}{2}\nu^{2}n_{0}^{2}\right\}
   \label{11.2}\\
    S_{\rm free}&=&\frac{1}{2}\summa_{{\sbf i},{\sbf j}}\int dt\varphi_{a}({\bf x}_{\sbf i}, t)M_{ab}
     ({\bf x}_{\sbf i}, t; {\bf x}_{\sbf j}, t)\varphi_{b}({\bf x}_{\sbf j}, t)\label{11.3}\\
      S_{\rm int}&=&\int dt\sum_{\sbf{i}}{\cal L}_{\rm int}(\varphi_{1}({\bf x}_{\sbf i},t),\varphi_{2}({\bf x}_{\sbf i},t))
       \label{11.4}
        \eea
        \be
        \ba
       {\cal L}_{\rm int}(\varphi_{1}({\bf x}_{\sbf i},t),\varphi_{2}({\bf x}_{\sbf i},t))= v_{3}[\varphi_{1}({\bf x}_{\sbf i}, t)
        \varphi_{2}^{2}({\bf x}_{\sbf i}, t)+\varphi_{1}^{3}({\bf x}_{\sbf i}, t)]\nonumber\\
         +v_{4}[\varphi_{1}^{4}({\bf x}_{\sbf i}, t)+
         \varphi_{2}^{4}({\bf x}_{\sbf i}, t)+2\varphi_{1}^{2}({\bf x}_{\sbf i}, t)\varphi_{2}^{2}({\bf x}_{\sbf i}, t)]
         \equiv{\cal L}_{3}+{\cal L}_{4}
         \label{11.5}
         \ea
         \ee
where $2\times2$ matrix $M_{ab}$ is given by  Eqs. \re{1.1911}, \re{1.191} ,
  $v_{3}=-U\sqrt{\nu n_0/2},\quad v_{4}=-U/8$.

The perturbative framework is based on the propagator  ${G}_{ab}(k,\omega)$ given in (\ref{1.20}). The effective potential  ${\cal{V}}$ can be evaluated  by the Eq. (\ref{1.9}),
where  the only connected, irreducible  diagrams in the partition function $Z=\int{\cal D}\varphi_{1}{\cal D}\varphi_{2}\exp{(iS(\varphi_{1},\varphi_{2}))}$ should be taken into account. The grand thermodynamic potential i.e. free energy, $\Omega({n}_0,\mu)$,
corresponds to the minimum of $\calv ( n_0,\mu)$, such that $ n_0$ is a
solution of the equation $\partial\calv ( n_0,\mu)/\partial  n_0=0$   \ci{andersen}.
Now using (\ref{11.2})-(\ref{11.4}) and making expansion by  ${\cal L}_{\rm int}$ one can represent $Z$ as follows:
\be
\ba
 Z=e^{iS_{0}}\int{\cal D}\varphi_{1}{\cal D}\varphi_{2}
  e^{iS_{\rm free}+iS_{\rm int}}\label{12.1}\\
   =e^{iS_{0}}\int{\cal D}\varphi_{1}{\cal D}\varphi_{2}e^{\frac{i}{2}
    \varphi_{a}M_{ab}\varphi_{b}}\Bigg\{1+i\sum_{\sbf{i}}\int dt {\cal L}_{\rm int}
     (\varphi_{1}({\bf x}_{\sbf i}, t),\varphi_{2}({\bf x}_{\sbf i}, t))\nonumber\\
      +\dsfrac{i^{2}}{2}\sum_{{\sbf i},{\sbf j}}\int dt dt^{\prime}{\cal L}_{\rm int}
       (\varphi_{1}({\bf x}_{\sbf i}, t),\varphi_{2}({\bf x}_{\sbf i}, t)){\cal L}_{\rm int}
        (\varphi_{1}({\bf x}_{\sbf j}, t),\varphi_{2}({\bf x}_{\sbf j}, t))\Bigg\}\nonumber\\
      =\dsfrac{e^{iS_{0}}}{\sqrt{{\rm Det}{G}}}\Bigg\{1+i\sum_{\sbf{i}}\langle{\cal L}_{\rm int}
       (\varphi_{1}({\bf x}_{\sbf i}, t),\varphi_{2}({\bf x}_{\sbf i}, t))\rangle_{0}dt\nonumber\\
        +\dsfrac{i^{2}}{2}\sum_{{\sbf i},{\sbf j}}\int dt dt^{\prime}\langle{\cal L}_{\rm int}
         (\varphi_{1}({\bf x}_{\sbf i}, t),\varphi_{2}({\bf x}_{\sbf i}, t)){\cal L}_{\rm int}
          (\varphi_{1}({\bf x}_{\sbf j}, t),\varphi_{2}({\bf x}_{\sbf j}, t))\rangle_{0}\Bigg\}
          \label{12.2}
           \ea
           \ee
where we introduced the following notation
\be
 \langle \hat{A}(\varphi_{a}({\bf x}_{\sbf i}, t),\varphi_{b}({\bf x}_{\sbf i}, t))\rangle_{0}=\left.
  \hat{A}\left(\frac{\delta}{i\delta j_{a}({\bf x}_{\sbf i}, t)},\frac{\delta}{i
   \delta j_{b}({\bf x}_{\sbf i}, t)}\right)e^{\displaystyle{-\frac{i}{2}j_{\alpha}{G}_
   {\alpha \beta}j_{\beta}}}
    \right|_{j=0},
    \label{12.3}
     \ee
suppressing the
 summation and integration signs over lattice sites
 and times $t$ and $t'$
in quandratic forms, for brevity.

The classical contribution to ${\cal{V}}$ is given by      factor $\exp{(iS_{0})}$ in (\ref{12.2})
\bea
 {\cal{V}}_0&=&\frac{i}{T}\ln{e^{iS_{0}}}=\frac{N_s \nu  n_0}{2}[{U\nu  n_0}-2 (\mu+Jz_0)]\label{13.1}
    \eea
The one-loop contribution to the thermodynamic potential - ${\cal{V}}_{1L}$, can be obtained
by using the free part of the action (\ref{11.3})     in (\ref{12.2}), neglecting
interaction terms:

\be
 {\cal{V}}_{1L}=\frac{i}{2T}\Tr \ln {\rm Det} \hat{M}=
 \frac{i}{2}\sum_{{\bfq}}\int\frac{d\omega}{(2\pi)}
  \ln{{\rm Det}{M}(\omega,\bfq)}\label{13.2}
   \ee
where ${M}(\omega,\bfq)$ is given by (\ref{1.191}).
One notices that the frequency sum, and with it  the $\omega$ integration,
is divergent. In fact,  to evaluate the frequency sum such as
 $\ds\sum _{n=-\infty}^{n=\infty}\ln (a^2+\omega_{n}^2)$,
 with $\omega_n=2\pi n T$
one differentiates it with respect to $a$ and, after performing
the summation over $n$, integrates it over $a$.
This procedure gives an additional divergent constant which
may be removed by an additive renormalization of the energy \cite{KS}.
Therefore, in the case of optical lattices, where the momentum integration is performed within
a finite volume there is no additional ultraviolet divergency coming from $q$ integration,
 but there is an infinite constant coming from the frequency  summation \ci{kapustabellac}.
 This divergent constant
can be removed by subtraction from ${\cal{V}}$ the thermodynamic potential for the ideal gas \ci{haugset}:
\bea
{\cal{V}}_{1L}^{\rm ren}&=&{\cal{V}}_{1L}(U)-{\cal{V}}_{1L}(U=0)
=\dsfrac{1}{2}\sum_{{\sbf q}}\cE(\bq)-\left.\dsfrac{1}{2}
\sum_{{\sbf q}}\cE(\bq)\right|_{U=0}\nonumber \\
&=&\dsfrac{1}{2}\summa_{{\sbf q}}[\cE(\bq)-\veps(\bq)+\mu+Jz_0],
\lab{om1lren}
\eea
where we have used Eqs.\re{1.181}, \re{1.21} and  performed integration  by $\omega$ using formulas given in the Appendix. Further, for simplicity, we shall  suppress
the superscript  in ${\cal{V}}_{1L}^{\rm ren}$.

The two-loop contributions to ${\cal{V}}$ are involved in second and third terms of (\ref{12.2}) as
\bea
 {\cal{V}}_{2L}=\frac{i}{T}\ln\Bigg\{1+i\sum_{\sbf{i}}\int\langle{\cal L}_{\rm int}
  \rangle_{0}dt
   +\frac{i^{2}}{2}\sum_{{\sbf i},{\sbf j}}\int dt dt^{\prime}\langle{\cal L}_{\rm int}
    {\cal L}_{\rm int}\rangle_{0}\Bigg\}\label{131.1}
     \eea
 The former includes ${\cal L}_{3}(\varphi_{1},\varphi_{2})$ which does not contribute to $Z$, since it is in
  odd power of $\varphi_{a}$, and hence:
 \bea
  \langle{\cal L}_{\rm int}\rangle_{0}=\langle{\cal L}_{4}\rangle_{0}
   =v_{4}\{\langle\varphi^{4}_{1}\rangle_{0}+\langle\varphi^{4}_{2}\rangle_{0}+2
  \langle\varphi^{2}_{1}\varphi^{2}_{2}\rangle_{0}\}
    \label{131.2}
     \eea
The same is true for
$\langle{\cal L}_{3}(\varphi_{a}({\bf x}_{\sbf i}, t)){\cal L}_{4}(\varphi_{b}({\bf x}_{\sbf i}, t))\rangle_{0}$ coming from the third term of (\ref{12.2}). As to the term
${\cal L}_{4}(\varphi_{a}({\bf x}_{\sbf i}, t)){\cal L}_{4}(\varphi_{a}({\bf x}_{\sbf i}, t))$ it also should be omitted since its contribution is beyond   two-loop corrections. Therefore
\bea
 {\cal{V}}_{2L}&=&\frac{i}{T}\ln\Bigg\{1+i\sum_{\sbf{i}}\langle{\cal L}_{4}(\varphi_{1}
  ({\bf x}_{\sbf i}, t),\varphi_{2}({\bf x}_{\sbf i}, t))\rangle_{0}\nonumber\\
   &&{}+\frac{i^{2}}{2}\sum_{{\sbf i},{\sbf j}}\int dt dt^{\prime}\langle{\cal L}_{3}(\varphi_{1}
   ({\bf x}_{\sbf i}, t),\varphi_{2}({\bf x}_{\sbf i}, t)){\cal L}_{3}(\varphi_{1}
    ({\bf x}_{\sbf j}, t),\varphi_{2}({\bf x}_{\sbf j}, t))\rangle_{0}\Bigg\}\label{14.1}
     \eea
.

The second term in the logarithm in Eq.(\ref{14.1}) can be expressed in terms of propagator as
\be
 \langle{\cal L}_{4}\rangle_{0}=v_{4}\Big[3(G_{11}^{2}(0)+G_{22}^{2}(0))
 +2G_{11}(0)G_{22}(0)+4G_{12}^{2}(0)\Big]\label{14.2}
    \ee
where we used the following abbreviation $x=(\bx,t)$ and the formulas
\bea
\langle\varphi_{a}(x)\varphi_{b}(x')\rangle_0&=&G_{ab}(x-x'),\nonumber\\
 \langle\varphi_{a}^{4}\rangle_0&=&3G_{aa}^{2}(0),\nonumber\\
  \langle\varphi_{1}^{2}\varphi_{2}^{2}\rangle_0&=&G_{11}(0)G_{22}(0)+2G_{12}^{2}(0)
  \label{14.3},
   \eea
 and introduced the notation
 \bea
  G_{ab}(0)=G_{ab}(x,x)=\left.\frac{1}{N_{s}}\sum_{{\sbf q}}\int\frac{d\omega}
   {(2\pi)} {G}_{ab}(\omega, \bq)e^{i\omega (t-t')}\right|_{t\rightarrow t'}
   \label{14.4}
    \eea
Note that $G_{12}(0)$ is the constant
 (see the Appendix)
\bea
 G_{12}(0)=\dsfrac{1}{N_s}\sum_{{\sbf q} }
\int \dsfrac{ d\omega }{2\pi}
\frac{\omega  }{\omega^2-\cE^2(\bq)+i\epsilon}
 = \dsfrac{i}{2N_s}\sum_{{\sbf q}}[1]  =-G_{21}(0)=\dsfrac{i}{2}
 \lab{g120}.
  \eea

The third term,
\bea
 \langle{\cal L}_{3}{\cal L}_{3}\rangle_0 &=&v_{3}^{2}\Big[\langle\varphi_{1}(x)
  \varphi_{2}^{2}(x)\varphi_{1}(y)\varphi_{2}^{2}(y)\rangle_0
+2\langle
    \varphi_{1}(x)\varphi_{2}^{2}(x)\varphi_{1}^{3}(y)\rangle_0+\langle\varphi_{1}^{3}(x)
     \varphi_{1}^{3}(y)\rangle_0\Big]\label{15.1}
      \eea
includes averages with six $\varphi_{a}$.
These may be evaluated via Wick theorem to yield
\be
\ba
\langle\varphi_{1}^{3}(x)\varphi_{1}^{3}(y)\rangle_{0}=
     6G_{11}^{3}(x,y)\\
 \langle\varphi_{1}^{3}(x)\varphi_{1}(y)  \varphi_{2}^{2}(y)
  \rangle_{0}=6G_{11}(x,y)G_{12}^{2}(x,y)\\

   \langle\varphi_{1}(x)\varphi_{2}^{2}(x)\varphi_{1}(y)
   \varphi_{2}^{2}(y)
   \rangle_{0}=
    4G_{22}(x,y)G_{12}(x,y)G_{21}(x,y)  +2G_{22}^{2}(x,y)G_{11}(x,y).
    \label{84}
      \ea
      \ee
We have omitted one-particle reducible diagrams such as
$G_{22}(0)G_{11}(x,y)G_{11}(0).$

Now, using (\ref{14.2}), (\ref{15.1})-(\ref{84}) in (\ref{14.1}), we finally obtain:
\bea
 {\cal{V}}_{2L}&=&\frac{UN_{s}}{8}\Big[3G_{11}^{2}(0)+3G_{22}^{2}(0)+
  2G_{11}(0)G_{22}(0)+4G_{12}^{2}(0)\Big]\nonumber\\
   &&{}-\frac{iU^2\nu  n_0}{2T}\sum_{{\sbf i},{\sbf j}}\int dt dt^{\prime}
\Big[G_{22}^{2}({\bf x}_{\sbf i},t;{\bf x}_{\sbf j},t')
    G_{11}({\bf x}_{\sbf i},t';{\bf x}_{\sbf j},t')\nonumber\\
     &&{}+3G_{11}^{3}({\bf x}_{\sbf i},t;{\bf x}_{\sbf j},t')+
6G_{11}({\bf x}_{\sbf i},t;{\bf x}_{\sbf j},t')
      G_{12}^{2}({\bf x}_{\sbf i},t;{\bf x}_{\sbf j},t')\nonumber\\
       &&{}+2G_{12}({\bf x}_{\sbf i},t;{\bf x}_{\sbf j},t')
G_{21}({\bf x}_{\sbf i},t;{\bf x}_{\sbf j},t')
G_{22}({\bf x}_{\sbf i},t;,{\bf x}_{\sbf j},t')\Big]\equiv
        {\cal{V}}_{2L}^{(1)}+{\cal{V}}_{2L}^{(2)}
        \lab{16.1}.
         \eea

%
%
The two-loop diagrams that contribute the thermodynamic potential are shown in Fig.~1.

\begin{figure}[h]
\begin{center}
\leavevmode
\includegraphics[width=0.6\textwidth]{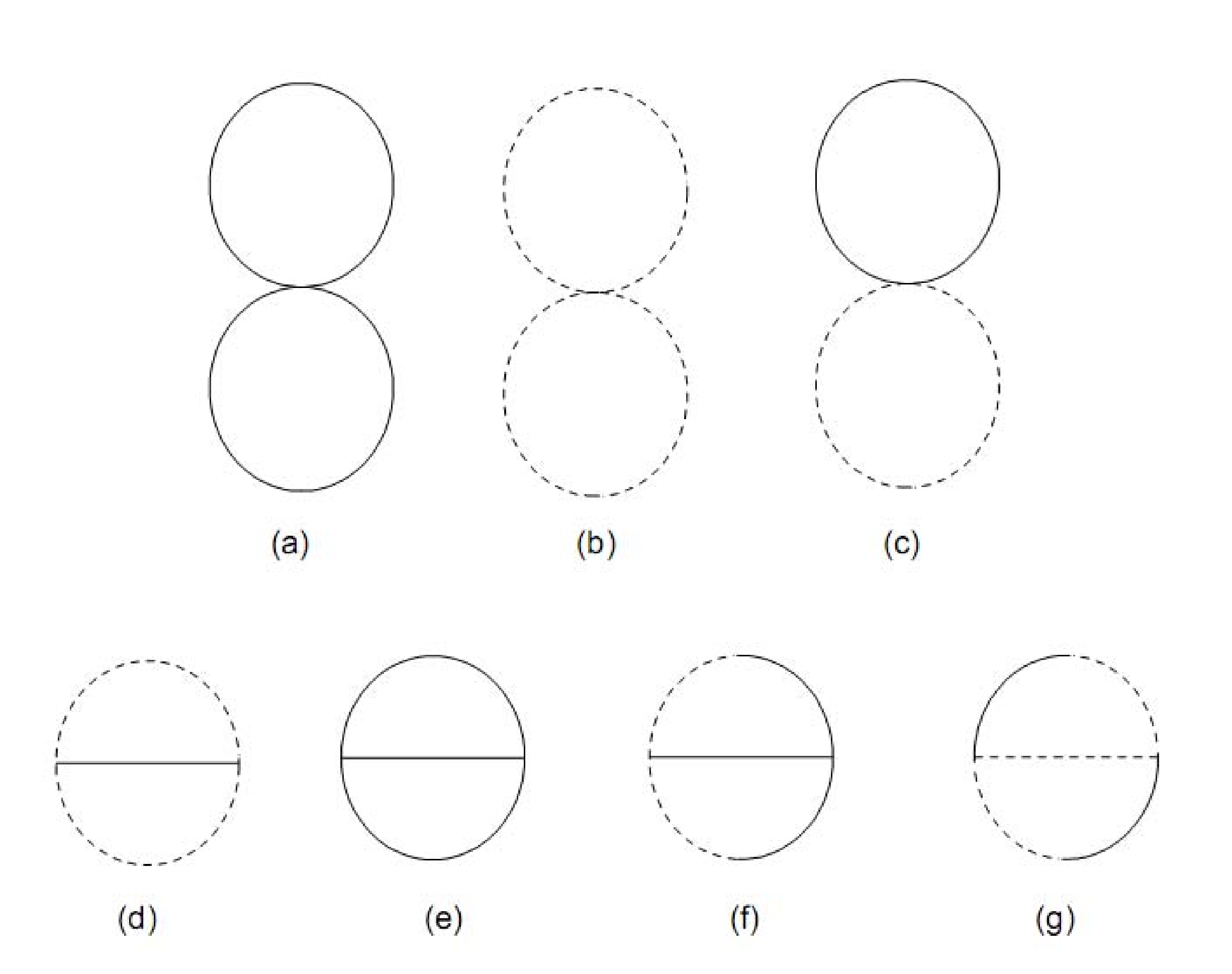}
\end{center}
\caption{Vacuum diagrams in a two-loop approximation. The solid and dashed lines  correspond to $G_{11}$ and $G_{22}$
 respectively, while the mixed line corresponds to $G_{12}$ (or $G_{21}$).}
\label{fig1}
\end{figure}



We now  pass to momentum space, and perform
integrations over energy variables $\omega$ to
obtain following analytic expression  (see Appendix):
\bea
 {\cal{V}}_{2L}^{(1)}( n_0,\mu)&=&\frac{U}{8}N_{s}\Big(3I_{10}^{2}+3I_{20}^{2}+
  2I_{10}I_{20}-1\Big),\label{17.1}\\
  {\cal{V}}_{2L}^{(2)}( n_0,\mu)&=&-\frac{U^{2}\nu  n_0}{8N_{s}}\Big(I_{1}+3I_{2}-6I_{3}
   +2I_{4}\Big)\label{17.2}.
    \eea
where following integrals are introduced
\be
\ba
 I_{10}( n_0,\mu)=\dsfrac{1}{N_{s}}\sum_{\sbq}\dsfrac{(-\tilmu+3U\nu n_0+
{\varepsilon}(\bq))}
  {2\cE(\bq)}=G_{22}(0),
  \nwl
   I_{20}( n_0,\mu)=\dsfrac{1}{N_{s}}\sum_{\sbq}\dsfrac{(-\tilmu+U\nu n_0+
   {\varepsilon}(\bq ))}
    {2\cE(\bq )}=G_{11}(0),\nwl
     I_{1}( n_0,\mu)=\sum_{\sbq_{1}\neq \sbq_{2}}\dsfrac{(-\tilmu+3U\nu n_0+{\varepsilon}(\bq _{1}))
      (-\tilmu+3U\nu n_0+\varepsilon(\bq _{2}))(-\tilmu+U\nu n_0+{\varepsilon}(\bq _{3}))}
       {\cE(\bq _{1})\cE(\bq _{2})\cE(\bq _{3})(\cE(\bq _{1})+\cE(\bq _{2})+\cE(\bq _{3}))},
            \nwl
     I_{2}( n_0,\mu)=\sum_{\sbq_{1}\neq \sbq_{2}}\dsfrac{(-\tilmu+U\nu n_0+{\varepsilon}(\bq _{1}))
      (-\tilmu+U\nu n_0+{\varepsilon}(\bq _{2}))(-\tilmu+U\nu n_0+{\varepsilon}(\bq _{3}))}
       {\cE(\bq _{1})\cE(\bq _{2})\cE(\bq _{3})(\cE(\bq _{1})+\cE(\bq _{2})+\cE(\bq _{3}))}
 ,
       \nwl
     I_{3}( n_0,\mu)=\sum_{\sbq_{1}\neq \sbq_{2}}\dsfrac{(-\tilmu+U\nu n_0+{\varepsilon}(\bq _{3}))}
      {\cE(\bq _{3})(\cE(\bq _{1})+\cE(\bq _{2})+\cE(\bq _{3}))}
  ,      \nwl
     I_{4}( n_0,\mu)=\sum_{\sbq_{1} \neq \sbq_{2}}\dsfrac{(-\tilmu+3U\nu n_0+{\varepsilon}(\bq _{3}))}
      {\cE(\bq _{3})(\cE(\bq _{1})+\cE(\bq _{2})+\cE(\bq _{3}))}
      \label{17.8}
   \ea
   \ee
         and $\cE(\bq )=\sqrt{(-\tilmu+3U\nu n_0+\varepsilon(\bq ))}
         \sqrt{(-\tilmu+U\nu n_0+\varepsilon(\bq ))}$,  $\tilmu=\mu-Jz_0$, $\bfq_{3}=
       \bfq_{1}-\bfq_{2}$.

Therefore the full effective potential in a two-loop approximation
is given by
\bea
 \calv(\mu, n_0)=\calv_{0}(\mu, n_0)+\calv_{1L}(\mu, n_0)+
 \calv_{2L}^{(1)}(\mu, n_0)
  +\calv_{2L}^{(2)}(\mu, n_0)\label{omtot}
   \eea
where
 $\calv_{0}$, $\calv_{1L}$, $\calv_{2L}^{(1)}$ , $\calv_{2L}^{(2)}$ are given by equations  \re{13.1}, \re{om1lren}, \re{17.1}, \re{17.2}
respectively.
Note that for the homogenous Bose gas.
Eqs. \re{17.1}-\re{omtot} were calculated before by
Braaten and Nieto \ci{braaten}.


\section{The condensate fraction in VPT}

To evaluate the condensate fraction
 $ n_0$
as an explicite
 function of $U/J$ and $\nu$ we shall use
following strategy referred as a variational perturbation theory\ci{KLVPT}:
\begin{enumerate}
\item
With fixed values of input parameters introduce an auxiliary parameter, loop counter,
 $\eta$ ($\eta$=1 at the end of calculations) to represent $\calv$ in Eq.
  \re{omtot}
 as:
\bea
   \calv(\mu,  n_0)=\calv_{0}(\mu,  n_0)+\eta\calv_{1L}(\mu, n_0)
    +\eta^{2}\calv _{2L}(\mu,  n_0)
    \label{omtoteta}
     \eea
     with $\calv _{2L}(\mu,  n_0)=\calv _{2L}^{(1)}(\mu,  n_0)+\calv _{2L}^{(2)}(\mu,  n_0)$
\item
Impose the  extremalization condition:
\bea
  \frac{\partial \calv(\mu,  n_0)}{\partial  n_0}=O(\eta^{3})
  \lab{a78}
   \eea
 and solve this equation with respect to $ n_0$. Let the solution of the equation is
 $ \bar{n}_0(\mu)$.
 Clearly the latter can be also represented in powers of $\eta$:
  \be
   \bar{n}_{0}(\mu)=n_{00}(\mu)+\eta n_{01}(\mu)+\eta^{2}n_{02}(\mu)
   \label{n0eta}
    \ee
    with
    \be
    \ba
    n_{01}(\mu)=-\dsfrac{\calv' _{1L}(\mu,  n_{00})}{\calv'' _{0}(\mu,  n_{00})}\\
    \\
    n_{02}(\mu)=-\dsfrac{ n_{01}^{2}(\mu)\calv''' _{0}(\mu,  n_{00})+
    2\calv' _{2L}(\mu,  n_{00})+2 n_{01}(\mu)\calv'' _{1L}(\mu,  n_{00})
     }{2\calv'' _{0}(\mu,  n_{00}) }
       \ea
    \ee
   where the prime denotes the derivative with respect
   to $n_0$, e.g.
    $\calv' _{1L}(\mu,  n_{00})=[\partial \calv _{1L}(\mu,  n_{0})/\partial n_0] \vert_{n_0=n_{00}}$ and $n_{00}$ is the solution to the equation
    $\calv' _{0}(\mu,  n_{0})=0$.

\item
 Inserting $ \bar{n}_0(\mu)$ back to the effective potential \re{omtoteta} determines the free energy
of the system $\Omega(\mu)=\calv( \bar{n}_0,\mu)$
\item
Introducing a variational parameter $M$ as
\be
\mu=M+r\eta
\lab{vst11}
\ee
with the abbreviation
\be
r=\frac{\mu-M}{\eta}
\lab{vst12}
\ee
 and inserting \re{vst11} into $\Omega (\mu)$ reexpand this $\Omega (M,\mu,r)$ in powers of $\eta$
 at fixed  $r$.
\item
Reinserting back $r$ from \re{vst12}
 optimize $\Omega (M,\mu)$ with respect to the variational parameter
$M$. This will fix $\mu$ as a function of the optimal $M=M_{opt}$, with
\be
M_{opt}=U\nu-Jz_0
\ee
\item
Finally, inserting this $\mu$ into   \re{n0eta} one finds an explicit expression
for $n_0$ as $n_0=n_0(U/J,\nu).$
\end{enumerate}

Below we consider each step in detail.
First, taking partial derivative with respect to $ n_0$ from Eq. \re{omtoteta}
one presents \re{a78} as
\begin{eqnarray}
\frac{\partial \calv( n_0,\mu)}{\partial  n_0}&=&\frac{\partial \calv_{0}( n_0,\mu)}{\partial n_0}
 +\eta\frac{\partial \calv_{1L}( n_0,\mu)}{\partial  n_0}+\eta^{2}\frac{\partial \calv_{2L}^{(1)}
  ( n_0,\mu)}{\partial  n_0}+\eta^2 \frac{\partial \calv_{2L}^{(2)}( n_0,\mu)}{\partial  n_0}=0
  \lab{domn0eq}\\
\frac{\partial \calv_{0}( n_0,\mu)}{\partial n_0}&=&-N_{s}\left[\nu\tilmu-U\nu^{2} n_0\right]\\
 \frac{\partial \calv_{1L}( n_0,\mu)}{\partial  n_0}&=&
 -\frac{U\nu}{2}\sum_{\sbq}\frac{(2\tilmu-3U\nu n_0-2\varepsilon(\bq)}{\cE(\bq)
 }\\
   \frac{\partial \calv_{2L}^{(1)}( n_0,\mu)}{\partial n_0}&=&\frac{U^{2}\nu}{4}\sum_{\sbq}\frac{(\tilmu-\varepsilon(\bq)
    )\left[(\tilmu-4U\nu n_0-\varepsilon(\bq))
I_{10}( n_0,\mu)-(\tilmu-\varepsilon(\bq))
    I_{20}( n_0,\mu)\right]}{\cE^{3}(\bq)}
  \nonumber\\
     {}&&
         \label{dev}
     \end{eqnarray}
     where following relations are used
    \bea
 \frac{\partial \cE(\bf q)}{\partial  n_0}&=&-\frac{U\nu}{\cE(\bq)}
(2\mu-3U\nu n_0-2\varepsilon(\bq)+2Jz_{0})\\
  \frac{\partial I_{10}}{\partial n_0}&=&\frac{U\nu}{2N_{s}}\sum_{\sbq}\frac{(\mu-\varepsilon(\bq)+Jz_{0})
   (\mu-3U\nu  n_0-\varepsilon(\bq)+Jz_{0})}{\cE^{3}(\bq)}
   \lab{aaa66}\\
    \frac{\partial I_{20}}{\partial n_0}&=&-\frac{U\nu}{2N_{s}}\sum_{\sbq}\frac{(\mu-\varepsilon(\bq)+Jz_{0})
     (\mu-U\nu  n_0-\varepsilon(\bq)+Jz_{0})}{\cE^{3}(\bq)}\label{a87}
      \eea
      In Eqs. \re{dev} $\partial \calv_{2L}^{(2)}/\partial  n_0 $ has a long expression
and will be given later. Solving Eq. \re{domn0eq} iteratively gives Eq. \re{n0eta}
with
\be
\ba
 n_{00}(\mu)=\displaystyle{\frac{\mu+Jz_{0}}{\nu U}},\nonumber\\
   n_{01}(\mu)=-\dsfrac{1}{2\nu}(3I_{20}(\mu)+I_{10}(\mu))=-\dsfrac{1}{2N_s\nu}
\sum_{\sbq}
  \dsfrac{(\mu+Jz_0+2\veps(\bq))}{2\cE_\mu(\bq)},
  \nonumber\\
     n_{02}(\mu)= \left.-\dsfrac{1}{N_{s}U\nu^{2}}
    \dsfrac{
     \partial\Omega_{2l}^{(2)}(n_{0},\mu)
     }
      {
      \partial n_{0}}
      \right|_{\displaystyle{n_{0}=n_{00}}}+\dsfrac{1}{2N_s\nu}\sum_\sbq \left[
      -\dsfrac{U\veps^2(\bq) (I_{10}(\mu)+I_{20}(\mu)    )  }{\cE_{\mu}^3(\bq)}\right.\nonumber\\
    \qquad\qquad +\left.\dsfrac{2U I_{20}(\mu)\veps(\bq) (\mu+Jz_0   )  }{\cE_{\mu}^3(\bq)}+
       \dsfrac{U (\mu+Jz_0   )^2 (I_{10}(\mu)-I_{20}(\mu)    )  }{\cE_{\mu}^3(\bq)}
      \right]
               \label{a88a},
               \ea
       \ee
      where
\bea
 I_{10}(\mu)&=&{\displaystyle\left.I_{10}( n_0,\mu)\right|_{ n_0=n_{00}}}=\frac{1}{2N_{s}}\sum_{\sbq}\frac{2\mu+2Jz_{0}+\varepsilon(\bq)}
  {\cE_{\mu}(\bq)},\nonumber\\
   I_{20}(\mu)&=&{\displaystyle \left.I_{20}( n_0,\mu)\right|_{ n_0=n_{00}}}=\frac{1}{2N_{s}}
   \sum_{\sbq}\frac{\varepsilon(\bq)}
   {\cE_{\mu}(\bq)}.
   \label{a89}
    \eea
  In this  step the
Goldstone boson dispersion is correctly achieved:
  \be
   {\cE_{\mu}(\bq)}=\sqrt{\varepsilon(\bq)}\sqrt{\varepsilon(\bq)+2\mu+2Jz_0}
   \lab{eemuk}
   \ee

Now inserting \re{n0eta}, \re{a88a} into \re{omtot} one gets $\Omega(\mu)$ as a function of $\mu$
as
\be
\ba
\Omega(\mu)={\cal V}(\mu,\bar{n}_0)=\Omega_0(\mu)+\eta\Omega_1(\mu)+\eta^2\Omega_2(\mu)\\
\\
\Omega_0(\mu)=-\dsfrac{N_s (\mu+Jz_0)^2}{2U},\\
\Omega_1(\mu)=\dsfrac{1}{2}\sum_\sbq [\cE_\mu(\bq)+\mu-\veps(\bq)+Jz_0],\\
\Omega_2(\mu)=\calv _{2L}^{(2)}(\mu ,n_{00}(\mu)  )+
\dsfrac{UN_s}{8}[2I_{10}^{2}(\mu)-4I_{10}(\mu)I_{20}(\mu)-6I_{20}^{2}(\mu)-1].
\lab{omtotmu}
\ea
\ee
Performed one more step of VPT we finally obtain $\mu$ as an explicit function
of the parameters $U,J,\nu$:
\be
\ba
\mu=\mu_0+\eta\mu_1+\eta^2\mu_2,\\
 \mu_0=U\nu-Jz_{0},\\
 \mu_1=\dsfrac{U}{2N_{s}}\sum_\sbq \dsfrac{\veps(\bq)+\cE_{0}(\bf q)}{\cE_{0}(\bf q)}=U\left(I_{20B}+\frac{1}{2}\right),\nwl
  \mu_2=\left.\dsfrac{U}{N_{s}}\dsfrac{\partial    \calv^{(2)}_{2L}(\mu)    }
    {    \partial\mu    }\right|_{\mu=\mu_0}+\dsfrac{U(I_{10B}-I_{20B})^2}{4\nu}
   \nwl
  ~~~~~~~ +\dsfrac{U^2(I_{10B}+I_{20B}-1)}{4N_s}\sum_\sbq
    \dsfrac{\veps^2(\bf q)}{\cE_{0}^{3}(\bf q)}
        \label{muu}.
\ea
\ee
and also the normal fraction, $n_1=1-\bar{n}_0$ as
\be
\ba
 n_{1}= n_{1}^{1L}+ n_{1}^{2L},\\
 n_{1}^{1L}=\normalsize
 \dsfrac{1}{2\nu N_{s}}\sum_{\sbq}\left[
\normalsize \dsfrac{\varepsilon(\bq)
 +U\nu}{\cE_{0}(\bf q)}-1\right],
 \lab{n1l}
 \ea
 \ee
 \be
\ba
      n_{1}^{2L}=\left.\dsfrac{1}{N_{s}U\nu^{2}}\dsfrac{\partial\calv_{2L}^{(2)}}
      {\partial  n_0}\right|_{n_0=n_{00}}
    -\left.\dsfrac{1}{\nu N_{s}}\frac{\partial\calv_{2L}^{(2)}}{\partial\mu}\right|_{\mu=\mu_0}

    -    \dsfrac{(I_{10B}-I_{20B})^2}{4\nu^2}
    \nwl
  ~~~~~~~  -\dsfrac{U}{4N_s\nu}\sum_\sbq\dsfrac{\left[(I_{10B}-I_{20B})(2U^2\nu^2-\veps^2(\bq))
    +U\nu\veps(\bq)(2I_{20B}-1)\right]}{\cE_{0}^{3}(\bf q)}
         \lab{a111}
     \ea
     \ee
In Eqs. \re{muu}, \re{a111}  $\cE_0(\bf q)$, $I_{10B}$ and  $I_{20B}$  are  given
 by
 \be
 \ba
 \cE_{0}({\bf q})=\sqrt{\varepsilon({\bf q})}\sqrt{\varepsilon(\bfq)+2U\nu},\nwl
  I_{10B}=\dsfrac{1}{2N_{s}}\sum_{\sbq}\dsfrac{2U\nu+\varepsilon(\bq)}{\cE_{0}(\bfq)},\nwl
   I_{20B}=\dsfrac{1}{2N_{s}}\sum_{\sbq}\dsfrac{\varepsilon(\bfq)}{\cE_{0}(\bfq)}
   \label{e0k}.
    \ea
    \ee

Now we compare present approximation with Gutzwiller's.
\begin{itemize}
   \item In Gutzwiller approach the phonon dispersion for small $\vec{q}$ is quadraric in wave number \ci{Krauth}
   rather than linear    given in present approximation
   by Eq. \re{eemuk}.\\
   \item As it is seen from Eq.s \re{n1l} and \re{a111} in Bogoliubov type approximations
   the uncondensed particles have momentum distribution $n_q=<{a}^{\dagger}_{q}{a}_{q}>$
varying as $q^{-4}$ for large momentum \ci{yukannals}, while in  Gutzwiller approach this distribution
is independent of  $\vec{q}$ \ci{Krauth}.
 \end{itemize}

\section{ground state energy}
The ground state energy of the system at zero temperature can be determined
as
\bea
 E=\Omega(\mu)+\mu N
  \lab{a112},
  \eea
where $\Omega(\mu)$ in Eq. \re{ommu}
 can be rewritten as follows
\be
\ba
 \Omega(U,J,\nu)=\Omega_0(U,J,\nu)+\Omega_1 (U,J,\nu)+\Omega_2(U,J,\nu),\nwl
 \Omega_0(U,J,\nu)=-\dsfrac{UN_s\nu^2}{2},\quad
 \Omega_1(U,J,\nu)=\dsfrac{1}{2}\sum_\sbq [\cE_0(\bq)-\veps(\bq)]+N_s \nu
\left(\frac{U}{2}-\mu_1\right),\nwl
  \Omega_2(U,J,\nu)=\Omega_{2L}^{(2)}(U,J,\nu)+\dsfrac{UN_s\left(
2I_{10B}^{2}-4I_{10B}I_{20B}
 -6I_{20B}^{2}-1
 \right)}{8}
 +\dsfrac{N_s(\mu_1^{2}-2U\nu\mu_2)}{2U}
       \lab{ommu}
      \ea
      \ee
     where $\Omega_{2L}^{(2)}$ is given by
\bea
&&\!\!\!\!\!\!\!\!\!\!\!\!\!\!\!\!\!\!\!\! \!\!\!\!\!\!\Omega_{2L}^{(2)}(U,J,\nu)=\calv_{2L}^{(2)}(n_{0}=1,\mu=\mu_{0})\nonumber \nwl
&&\!\!\!\!\!\!\!\!\!\!\!=
  -\dsfrac{NU^{2}}{4N_{s}^{2}}\summa_{\sbq_{1},\sbq_{2}}\Bigg[\dsfrac{
  U\varepsilon_{3}\nu(\varepsilon_{1}+\varepsilon_{2}+2U\nu) }{\cE_{0}(1)\cE_{0}(2)\cE_{0}(3)\cE_{0T}}
  +\dsfrac{2\varepsilon_{1}\varepsilon_{2}\varepsilon_{3}-2\cE_{0}(1)\cE_{0}(2)
   (\varepsilon_{3}+U\nu)}
    {\cE_{0}(1)\cE_{0}(2)\cE_{0}(3)\cE_{0T}}\Bigg],
     \lab{a116}
         \eea
     with $\cE_0(\bf q)$ given in \re{e0k}, and
     $\varepsilon_{1}\equiv \varepsilon_{\sbq_1} $, $\cE_{0}(1)\equiv \cE_0(\bq_1)$,
     $\cE_{0T}\equiv \cE_{0}(1)+\cE_{0}(2)+\cE_{0}(3)$.

After some algebraic manipulations one obtains for the energy per particle
$E/N$ following expression
\bea \dsfrac{E}{N}&=&\dsfrac{U(4\nu^2+4\nu-1)}{8\nu}
  +\dsfrac{\mu_{1}^{2}}{2U\nu}+\dsfrac{U(I_{10B}+I_{20B} )(I_{10B}-3I_{20B})    }{4\nu}
   +\dsfrac{\Omega_{2L}^{(2)}(U,J,\nu)}{N}
\nonumber\\
  ~~~~  &+&\dsfrac{1}{2N_{s}\nu}
\sum_{\sbq}[\cE_{0}(\bq) -\varepsilon(\bq)].
    \label{en}
     \eea
   Here the energy of an "ideal gas " (when $U=0$ in Bose-Hubbard Hamiltonian)
      has been subtracted .

 \section{Results and discussions}

 Firstly we discuss the condensate fraction, $ n_0$ vs $U/J$. In Fig.2a it is presented in
 one- and two-loop approximations, (dashed and solid curves respectively) for the filling factor
 $\nu=1$ and $D=3$.  It is seen that in the one loop approximation $ n_0$ can not reach zero within moderate
 values of $\uj$. More precisely $ n_0 [{\rm one~loop}]=0 $ at $\uj=81.2$ . On the other hand,
 two-loop contributions coming from the diagrams in Fig. 1 are too  large: quantum phase
 transition occurs at $\uj\simeq 6$. Unfortunately this is rather far from the experimental
 value: $ n_0=0$ at $\uj\simeq 29.34$ as pointed out in the  Introduction. It is seen from Fig.2b
 that in Gutzwiller approach $n_0$ reaches zero at $\uj\simeq 34.8$ \ci{Sheshadri}. Note that
 the similar behavior of $n_0$ vs $U/J$ with exactly the same $\kappa_{crit}$ has been found
 by Stoof et al. in decoupling approximation in  the second order perturbation theory \ci{Stoofmakola}.

\begin{figure}[h]
\begin{minipage}[h]{0.49\linewidth}
\center{\includegraphics[width=1.2\linewidth]{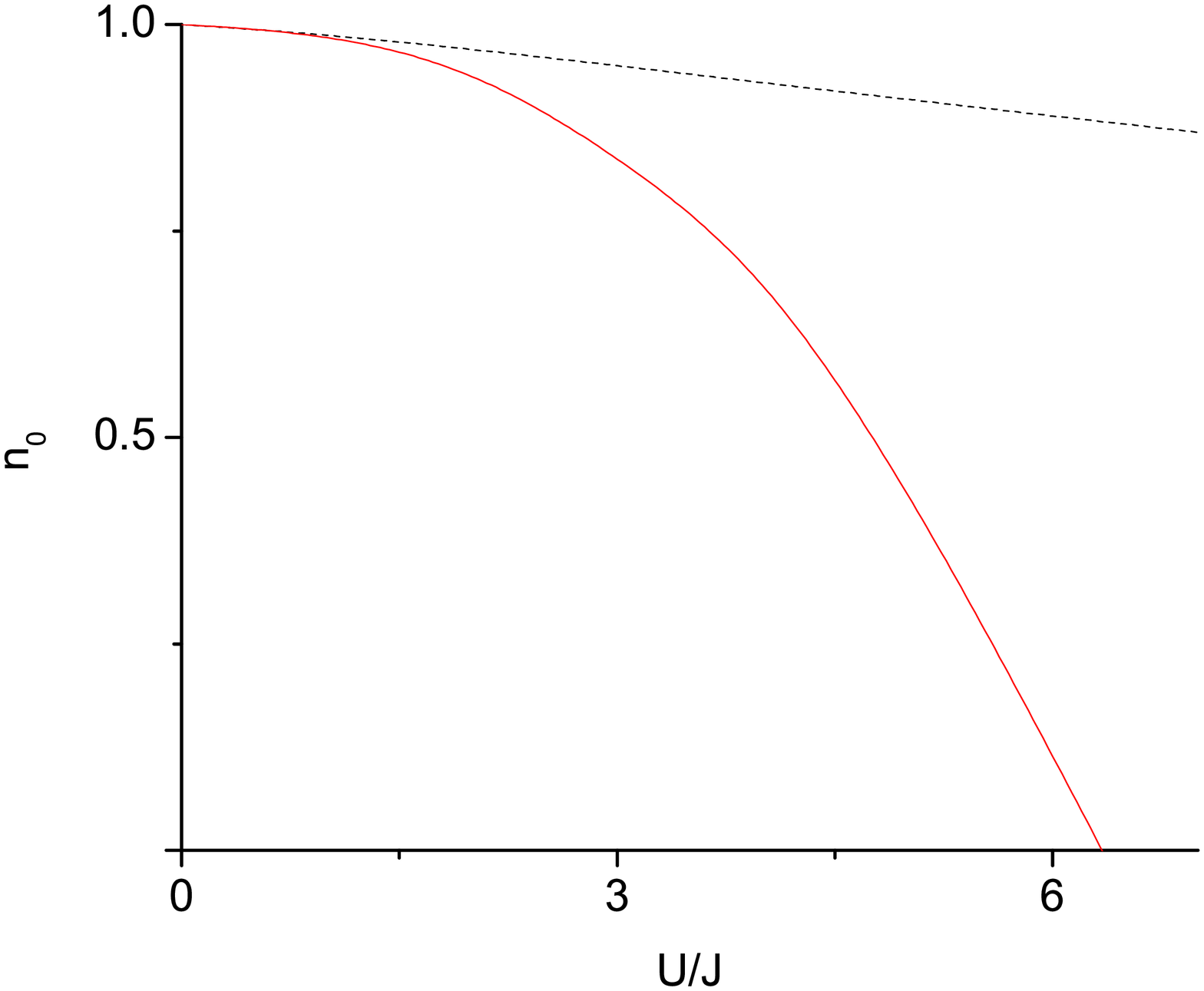} \\ a)}
\end{minipage}
\hfill
\begin{minipage}[h]{0.49\linewidth}
\center{\includegraphics[width=1.2\linewidth]{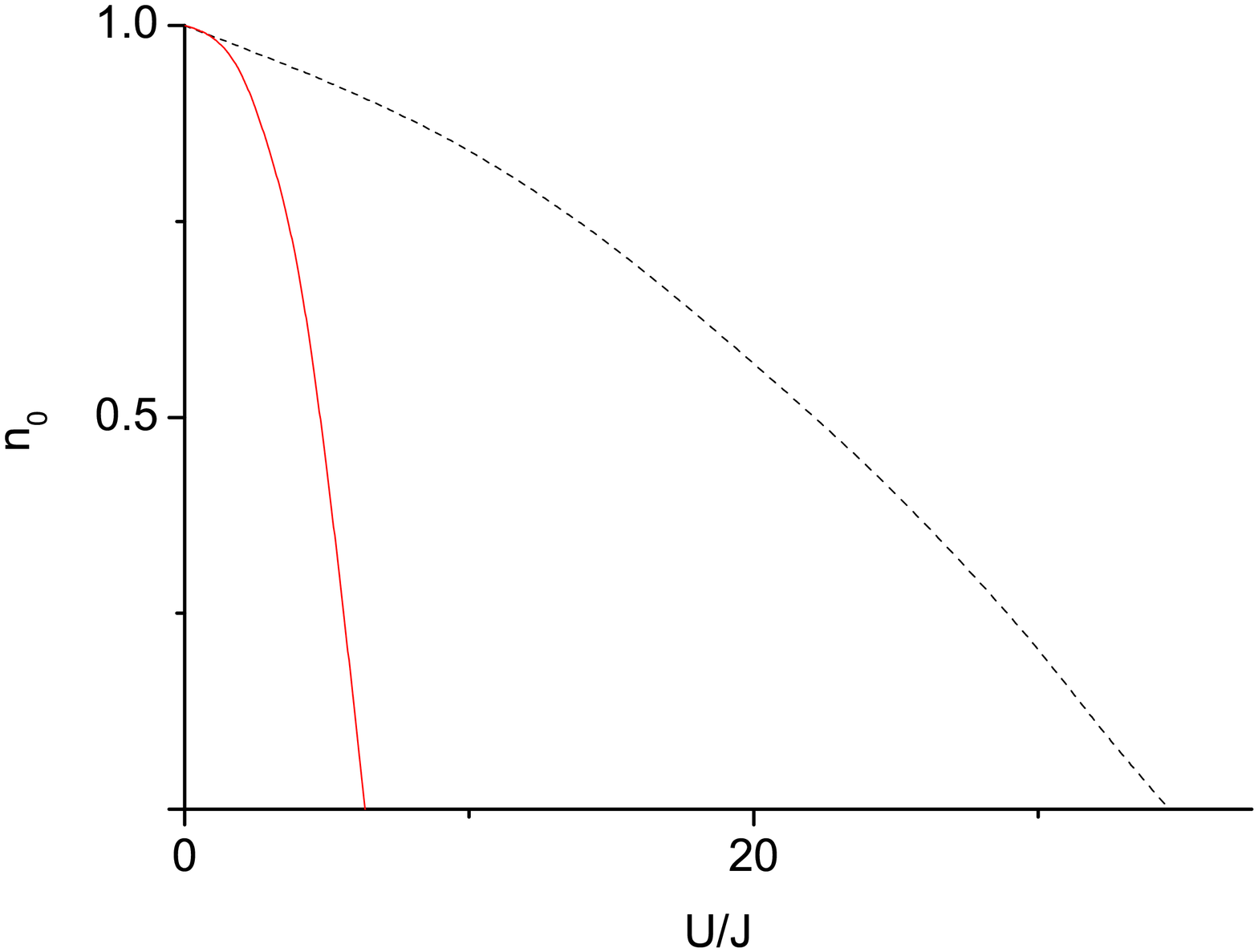} \\ b)}
\end{minipage}
\caption{ (Color online) The superfluid fraction $n_0$ as a function of  $U/J$ for $\nu=1$, $D=3$.
a)In one (dashed line) and two loop approximations  (solid line); b)
Here the dashed  line was  obtained in Gutzwiller approache
 while the solid line in the  present one.}
\label{fig:2}
\end{figure}

\begin{figure}[h]
\begin{minipage}[h]{0.49\linewidth}
\center{\includegraphics[width=1.2\linewidth]{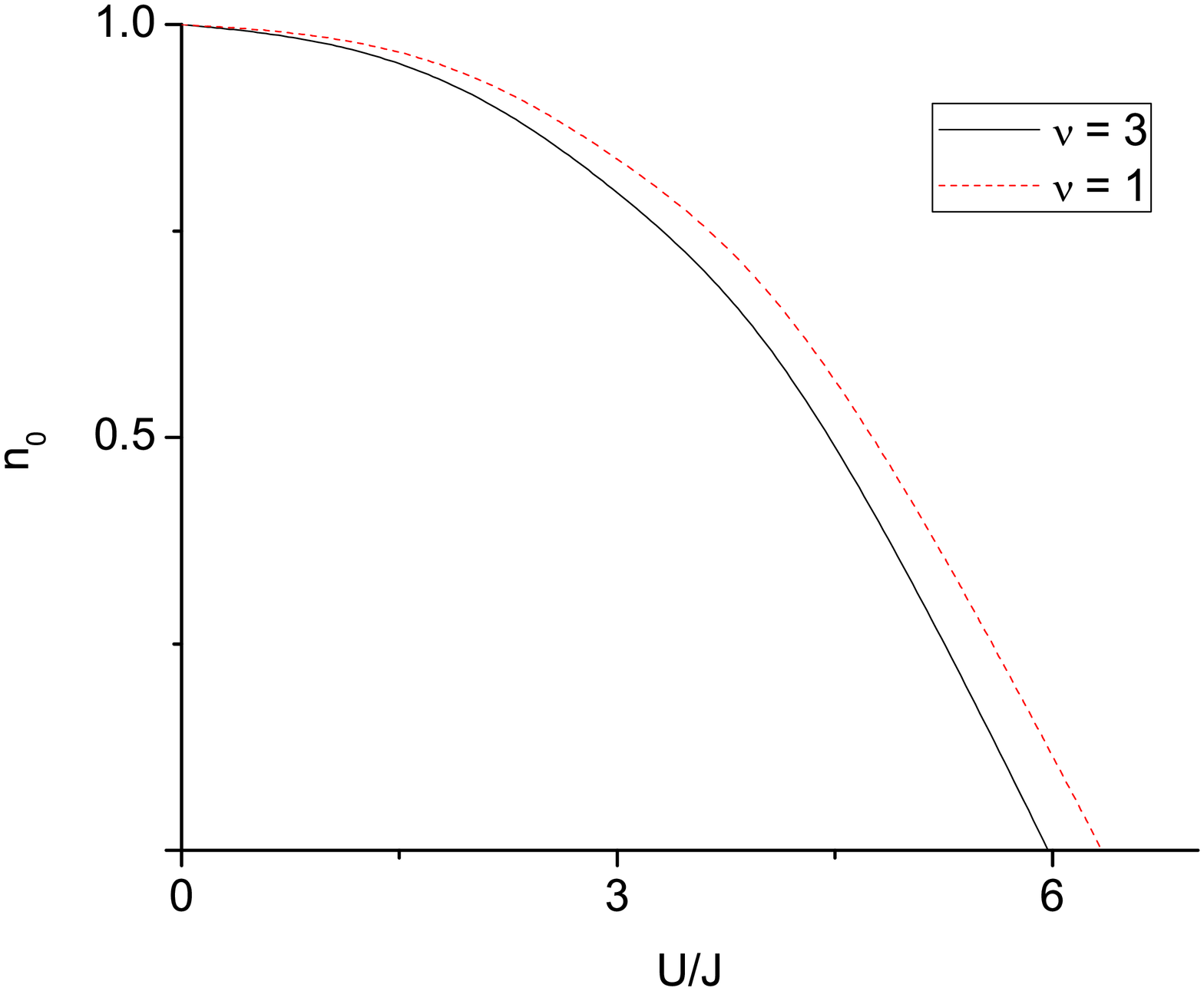} \\ a)}
\end{minipage}
\hfill
\begin{minipage}[h]{0.49\linewidth}
\center{\includegraphics[width=1.2\linewidth]{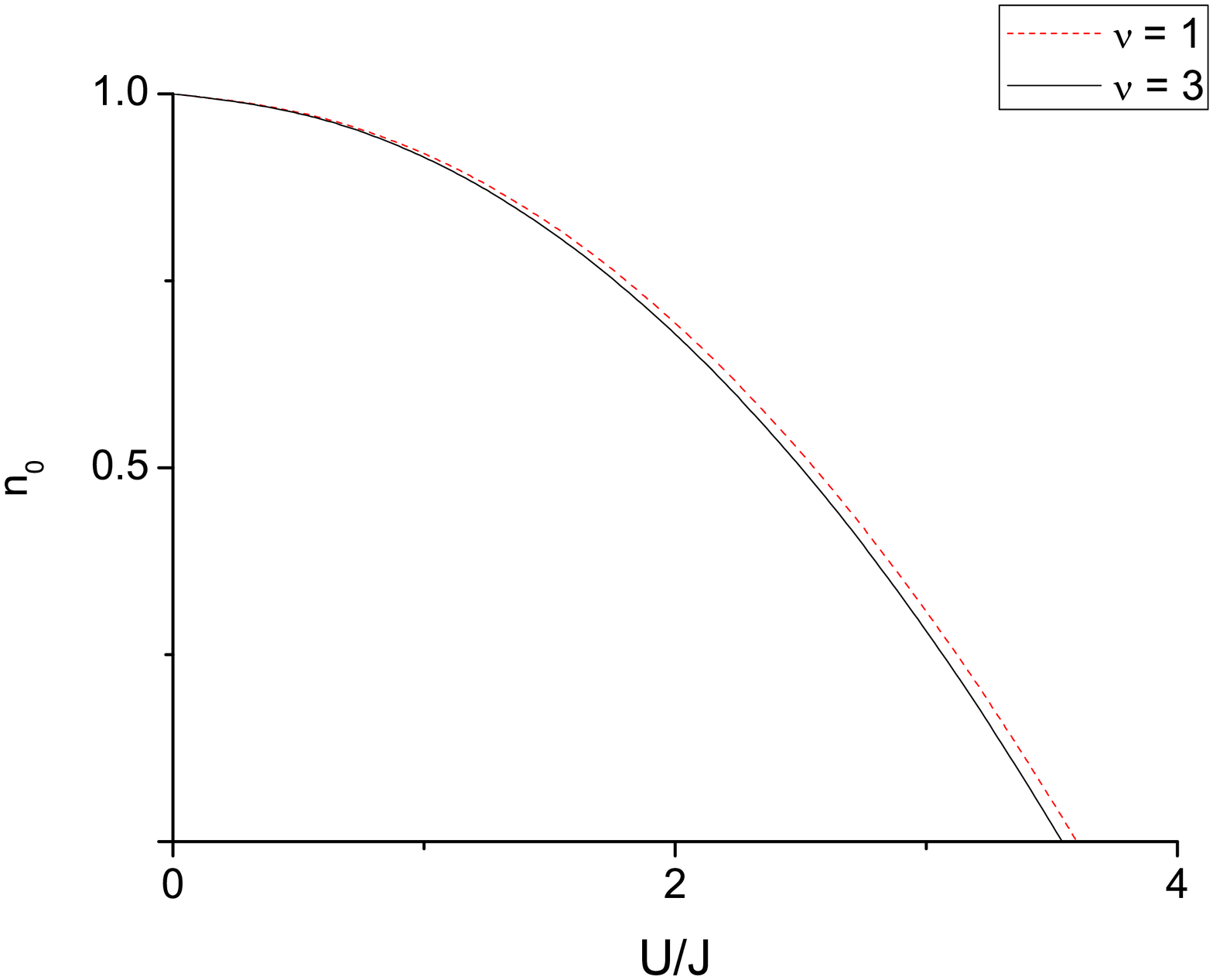} \\ b)}
\end{minipage}
\caption{(Color online) The superfluid fraction as a function of  $U/J$ for $\nu=1$ (dashed line) and
   $\nu=3$ (solid line) for a) $D=3$  and  b) $D=1$  in a two-loop approximation.}
\label{fig:3}
\end{figure}

\begin{figure}[h]
\begin{center}
\leavevmode
\includegraphics[width=0.6\textwidth]{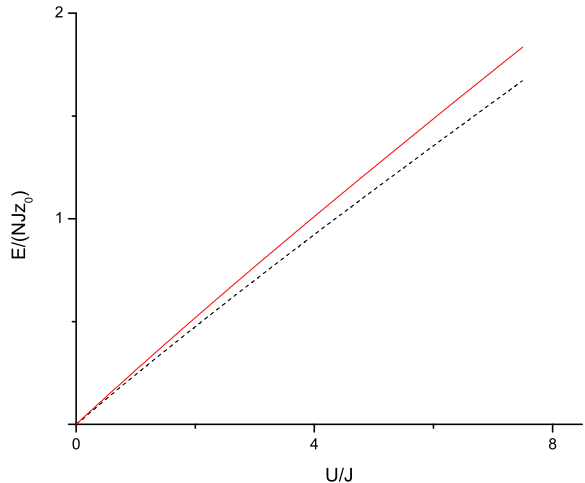}
\end{center}
\caption{ (Color online) The energy per atom in units $Jz_0$ in one (solid line)
  and two-loop (dashed line) approximations for $\nu=1$ for $D=3$.}
\label{fig4}
\end{figure}

\begin{figure}[h]
\begin{minipage}[h]{0.49\linewidth}
\center{\includegraphics[width=1.2\linewidth]{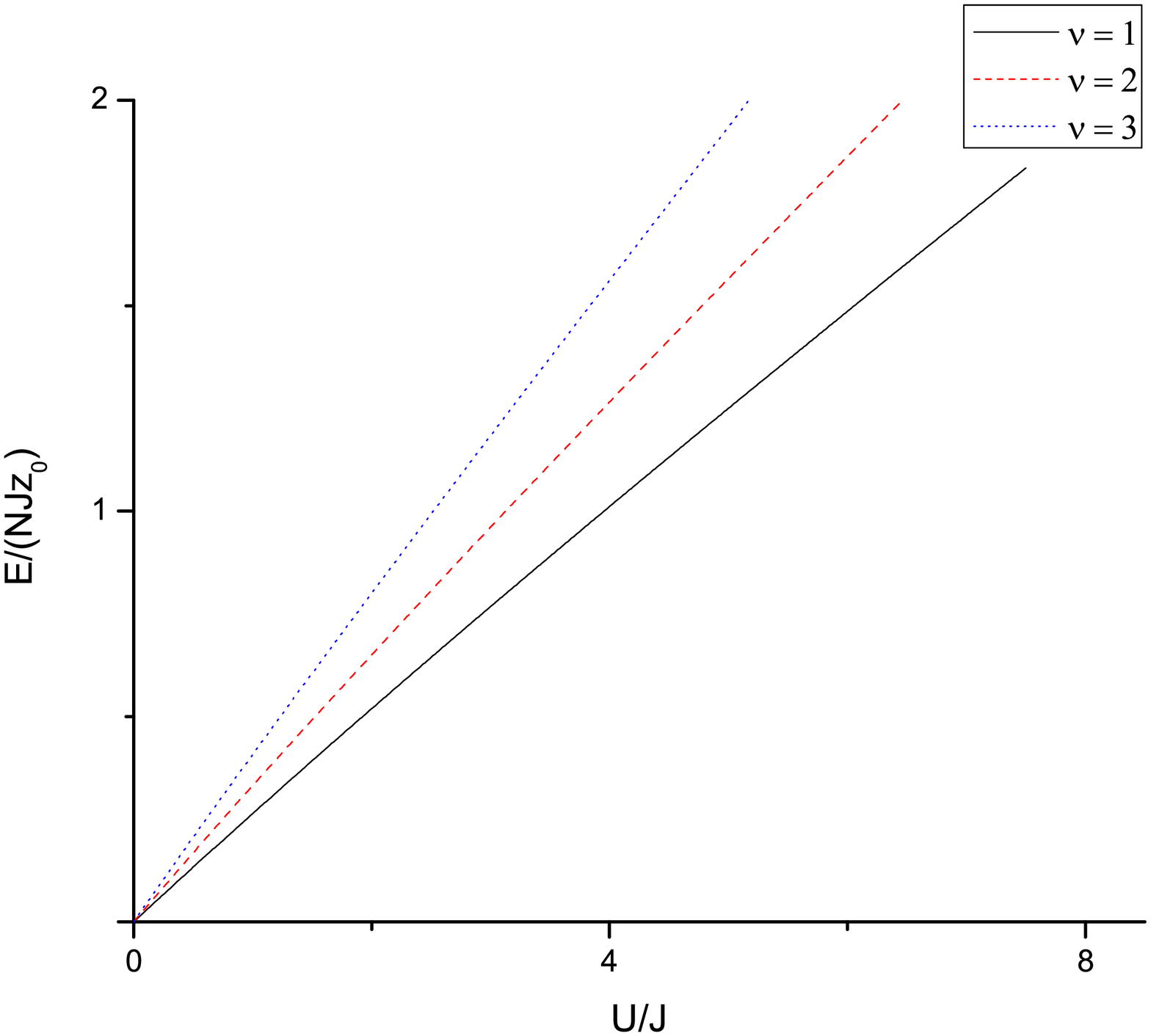} \\ a)}
\end{minipage}
\hfill
\begin{minipage}[h]{0.49\linewidth}
\center{\includegraphics[width=1.2\linewidth]{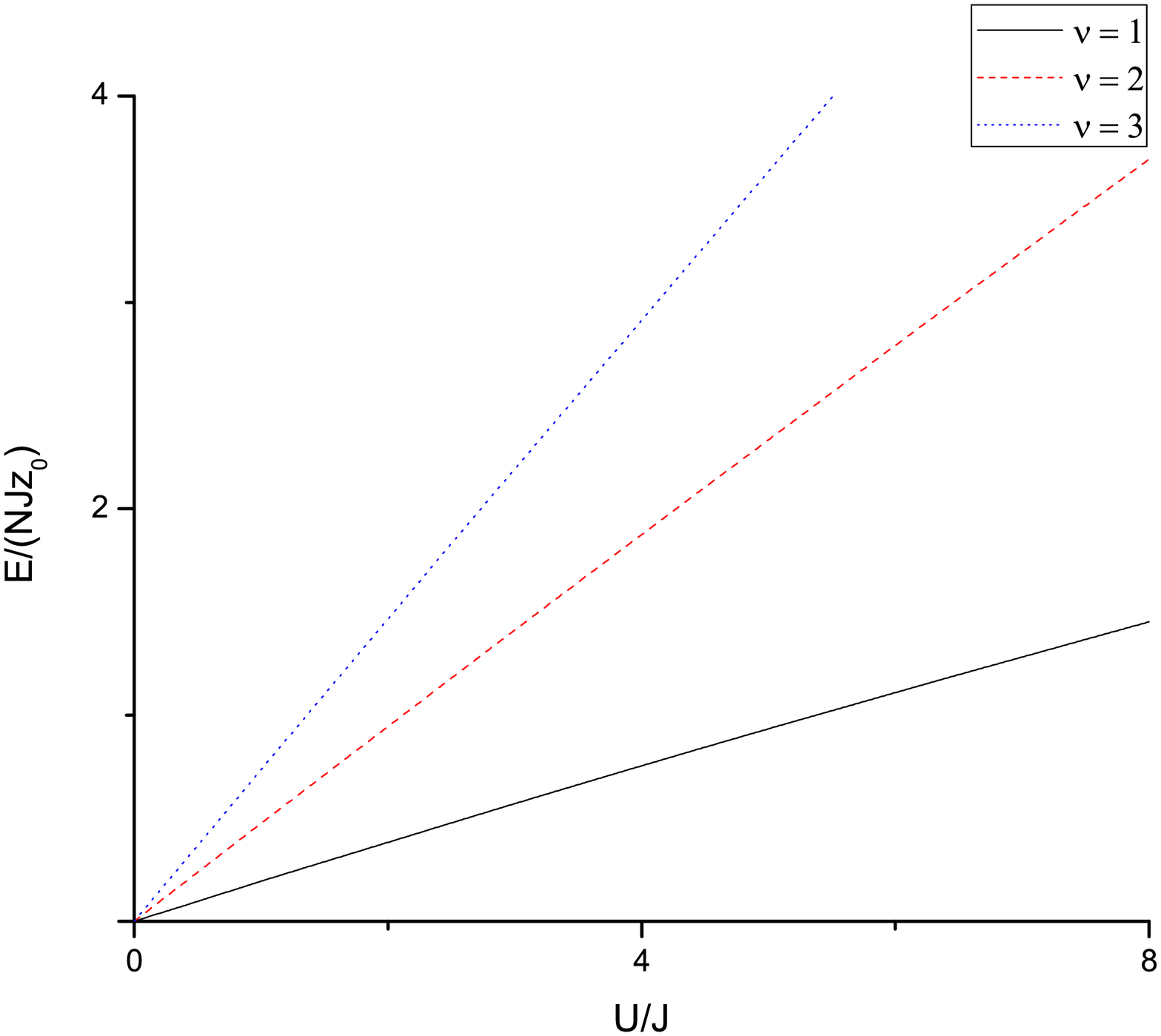} \\ b)}
\end{minipage}
\caption{(Color online) The energy per atom in units $Jz_0$ for various values of
   the filling parameter $\nu$ for a) $D=3$ and b) $D=1$.}
\label{fig:5}
\end{figure}

 The superfluid fraction for two values of $\nu$ ,  $\nu=1$ (dotted line) and $\nu=2$ (solid line) is shown
 in Fig. 3a and Fig. 3b for $D=3$ and $D=1$ respectively.
  It is seen that the critical value of $\uj$ as well as a whole $ n_0(\uj,\nu)$  are not
 so sensitive to the filling factor. The fact that the superfluid fraction does not crucially
 depend on $\nu$ has been observed also in Bogoliubov  \ci{stoofbook} as well as  HFB \ci{ourYUK} approximations.
 This is in contradiction with the prediction by Gutzwiller single site approximation
 \ci{Gutzwiller} where the dependence is rather strong:
 \be
 \kappa_{\rm crit}=z_0[\sqrt{\nu}+\sqrt{1+\nu} ]^2=2D[\sqrt{\nu}+\sqrt{1+\nu} ]^2
 \lab{gutz}
 \ee
  Note that, although  Eq. \re{gutz} gives a nice value for $\nu=1$,
  $\kappa_{\rm crit}=34.8$, it   can not be considered as an absolute truth
     since , besides it's drawbacks, outlined above, it  takes into account the lattice dimensionality in
     a rather simple way.

  On the other hand as it is seen from Fig. 3b, for $D=1$ the quantum phase transition,
  which , more strictly speaking, is  a Berezinskii- Kosterlitz- Thouless transition, occurs
  around $U/J=4$. This is in good agreement with Monte- Carlo predictions
  \ci{batrouni}. Similar results for $D=1$ have been obtained by Danshita and Naidon
  in their time - evolving block decimation (TEBD) method \ci{danshita}.
  However, note that TEBD method takes several days of computer calculations
  , while present approach does several minutes.
  In our calculations we used $N_s=60$, $N=\nu N_s$, that is we considered finite size
  systems. This explains the smoothness of $n_0 (U/J)$ in Figs. 3a and 3b.

  The ground state energy per particle $E/N$ vs $\uj$ in units $Jz_0$
  in one (solid line) and two (dashed line ) loops is presented in Fig. 4.
  It is normalized such that the appropriate energy for the
  ideal case ($U=0$ in the Bose-Hubbard Hamiltonian)   is set to zero.
  It is seen that quantum corrections due to diagrams in Fig.1 are not
  significant
  for small $U/J<1$.
  The dependence of $E/N$ on filling factor $\nu$
   is
  illustrated  in Figs. 5a, 5b.  It is seen that
  $E$ is more sensitive to $\nu$ than $ n_0$ due to the leading term
  ( the first term in Eq. \re{en}) depending on $\nu$ explicitly.


 \section{Summary and conclusions}

 We have developed a field theoretical approach in terms of path integral formalism  to  calculate the second-order quantum corrections to the energy density as well as to
 the superfluid fraction in cubic optical lattices.
Instead of using the standard formalism
 with complex field operatorsof condensed-matter literature,
we find it more convenient to use
 two real fields. The thermodynamics of the system is deduced
 from the effective potential $\calv$, whose minimum gives free energy $\Omega$.

 The  superfluid fraction, $ n_0$ ,  goes to zero at $U/J\sim 6$ for $\nu=1,2,3$, and this is
interpreted as a quantum phase transition from the superfluid
to the Mott insulator phase.
 For $D=1$, we have found a good description
of the  transition.  Unfortunately, for $D=2$ and   $D=3$
 the critical values for the parameters are rather far from the experiment:
 $\kappa^{\rm exp}_{\rm crit}(D=2)=16.8$ and   $\kappa^{\rm exp}_{\rm crit}(D=3)=29.34,$  for $\nu=1$.  It appears
that a more relaible value for $\kappa_{\rm crit}$
  for $D=2, 3$  can only be reached by
going beyond the present two-loop approximation.
 We expect that higher-order quantum corrections,  for example
 post-Gaussian approximation \ci{oursinggap,stev}, will
improve the situation, but they are hard to calculate.

Thus we have shown that going beyond the Bogoliubov approximation employed
by Stoof et al. \ci{Stoofmakola}, one
finds a
quantum
 phase transition from a superfluid to a
Mott
 insulator state. Within a two-loop approximation we have derived explicit expression for the ground state energy
of the optical lattice.

 \section*{Acknowledgments}
We acknowledge support of the Volkswagen Foundation. AR is also indebted to the DAAD for
partial support and to V. Yukalov and A. Pelster  for useful discussions.
\newpage
\section*{Appendix  }

In present work all the  calculations are carried out in real time. Loop integrals
are taken over real energies $\omega$ and over three dimensional quasimomentum $\vec{k}$
which pertains to the Brillouin zone $-\pi/a\leq k_\alpha\leq \pi/a.$ So, three or six
dimensional integrals, presenting in one or two-loop calculations are finite and may be
evaluated numerically by using Monte - Carlo methods.

The integrals over $\omega$ are evaluated using contour integration. Some energy integrals needed for one- and two-loop calculations can be easily evaluated directly by using residue formulas:

\be
\dsint _{-\infty}^{+\infty}\dsfrac{d\omega}{2\pi}\dsfrac{1}{(\omega^2-\cE^2+i\epsilon)}=-\dsfrac{i}{2\cE}
\lab{A1}
\ee
\be
\dsint _{-\infty}^{+\infty}\dsfrac{d\omega}{2\pi}\dsfrac{1}{(\omega^2-\cE^2+i\epsilon)^2}=\dsfrac{i}{4\cE^3}
\lab{A2}
\ee
\be
\dsint _{-\infty}^{+\infty}\dsfrac{d\omega}{2\pi}\dsfrac{\omega^2}{(\omega^2-\cE^2+i\epsilon)^2}=-\dsfrac{i}{4\cE}
\lab{A3}
\ee
\be
\dsint _{-\infty}^{+\infty}\dsint _{-\infty}^{+\infty}
\dsfrac{d\omega_1 d\omega_2}{4\pi^2}\dsfrac{1}
{[\omega_{1}^{2}-\cE_{1}^{2}+i\epsilon][\omega_{2}^{2}-\cE_{2}^{2}+i\epsilon][(\omega_{1}+\omega_2)^{2}-\cE_{3}^{2}+i\epsilon]         }=\dsfrac{1}{4\cE_1\cE_2\cE_3(\cE_1+\cE_2+\cE_3)}
\lab{A4}
\ee
\be
\dsint _{-\infty}^{+\infty}\dsint _{-\infty}^{+\infty}
\dsfrac{d\omega_1 d\omega_2}{4\pi^2}\dsfrac{\omega_1\omega_2}
{[\omega_{1}^{2}-\cE_{1}^{2}+i\epsilon][\omega_{2}^{2}-\cE_{2}^{2}+i\epsilon][(\omega_{1}+\omega_2)^{2}-\cE_{3}^{2}+i\epsilon]         }=\dsfrac{1}{4\cE_3(\cE_1+\cE_2+\cE_3)}
\lab{A5}
\ee
In the last two integrals $\cE_1\equiv \cE(\bq_1)$, $\cE_2\equiv \cE(\bq_2)$, and $
\cE_3\equiv \cE(\bq_1+\bq_2)$.

The integral
\be
I_{12}(\bq)=\dsint \dsfrac{d\omega}{2\pi}\dsfrac{i\omega}{(\omega^2-\cE^2(\bq)+i\epsilon)}
\lab{A6}
\ee
needed for $G_{12}(0)=-(i/N_s)\sum_{{\sbf q}} I_{12}(\bq)$ should be considered more carefully.To
evaluate it we use following formula given in the literature \ci{kapustabellac}
\be
\left.\dsfrac{1}{\beta}\ds\sum_{n=-\infty}^{\infty}\dsfrac{e^{i\eta\omega_n} (b+i\omega_n)}
{\omega_{n}^{2}+a^2}\right| _{\eta\rightarrow 0}=\dsfrac{1}{2}\left(
\dsfrac{b}{a}-1\right)+\dsfrac{b}{a(e^{\beta a}-1)}
\lab{A7}
\ee
where $\omega_n=2\pi n T$, $\beta=1/T$. The zero temperature limit, $T\rightarrow 0$, of \re{A7}
leads to
\be
I_{12}(\bq)=-\dsfrac{1}{2}
\lab{A8}
\ee
so that $G_{12}(0)=(i/2N_s)\sum_{{\bfq}} [1]$.
This constant
enters into the evaluation of the
constant
$n_1\sim \langle{\tilde\varphi}^{\ast}{\tilde\varphi}\rangle$, and produces
a term $-1$ in the square brackets
of Eq. \re{n1l}.
In a homogeneous Bose gas,
such a constant term can be ignored.
But here, on an optical lattice, it becomes significant,
so that
in the evaluation of trace log term in  Eq \re{13.2}, it must
be taken into account properly.
How to do that has been shown in the textbook
\cite{PI}. Strictly speaking, the integral
\be
L(\cE)=\dsint
\dsfrac{d\omega}{2\pi}\ln (\omega^2-\cE^2)
\lab{A9}
\ee
 appearing in the trace log
is divergent. To evaluate it, one may differentiate \re{A9} with respect to $\cE^2$:
\be
\dsfrac{\partial L(\cE)}{\partial \cE^2}=-\dsint \dsfrac{d\omega}{2\pi}\dsfrac{1}{(\omega^2-\cE^2)}
\lab{A10}
\ee
and use \re{A1} to obtain
\be
\dsfrac{\partial L(\cE)}{\partial \cE^2}=\dsfrac{i}{2\cE}
\lab{A11}
\ee
Integrating this once $\cE^2$ gives
\be
L(\cE)=\dsint
\dsfrac{d\omega}{2\pi}\ln (\omega^2-\cE^2)=i\cE+{\rm constant}.
\lab{A12}
\ee
Using the method of Ref.~\cite{PI}
we obtain the result of Section III where
the constant leads to a term
term $-1$ in  $n_1$ (see Eq. \ref{n1l}).
\newpage



\bb{99}
\bi{Morchrev} O. Morsch and M. Oberthaler Rev. Mod. Phys.
{\bf 78}, 179 (2006).
\bi{rous2003}  R. Raussendorf, D.E. Browne and H. J. Briegel, Phys. Rev. A {\bf 68}, 022312 (2003).
\bi{fisher} M. Fisher, P. B. Weichman, G. Grinstein and D. S. Fisher,
  Phys. Rev. B {\bf 40}, 546 (1989).
\bi{Greiner}M. Greiner, O. Mandel, T. Esslinger, T. W. Hansch and I. Bloch,
Nature {\bf 415}, 39 (2002).
\bi{svistun}B. Capogrosso-Sansone, N. V. Prokofiev, and B. V. Svistunov,
Phys. Rev. {\bf B} 75, 134302 (2007);\\
 S. Trotzky, L. Pollet, F. Gerbier, U. Schnorrberger, I. Bloch, N. V. Prokof'ev, B. Svistunov and M. Troyer, Nature Phys. {\bf 6}, 998-1004 (2010).
\bi{batrouni} G. G. Batrouni, V. Rousseau, R. T. Scalettar, M. Rigol,  A. Muramatsu, P. J. H. Denteneer, M. Troyer,  Phys. Rev. Lett. {\bf 89}, 117203 (2002).
\bi{Burger} S. Burger, F. S. Cataliotti, C. Fort, F. Minardi,  M. Inguscio, M. L. Chiofalo and M. P. Tosi, Phys. Rev. Lett. {\bf 86}, 4447 (2001).
\bi{Pelster}F. E. A. dos Santos and A. Pelster, Phys. Rev. A {\bf 79}, 013614 (2009).
\bi{stoofbook} H. T. C. Stoof, K. B. Gubbels and D.B.M. Dickerscheid
{\it Ultracold Quantum Fields} (Springer, 2009).
\bi{Gutzwiller} M. G. Gutzwiller, Phys. Rev. Lett. {\bf 10}, 159 (1963)
 \bi{RJ} D. S. Rokhsar and B. G. Kotliar Phys. Rev. B {\bf 44}, 10328 (1991) ;
\\D. Jaksch, C. Bruder, et al. Phys. Rev. Lett. {\bf 81}, 3108 (1998).
\bi{Krauth}W. Krauth , M. Caffarel,and J. Bouchaud Phys. Rev. B {\bf 45}, 3137 (1992)
\bi{Yukalovobsor} V. I. Yukalov, Laser Physics  {\bf 19}, 1 (2009).
\bi{vezzani} P. Buonsante and A. Vezzani, Phys. Rev. A {\bf 70}, 033608 (2004).
\bi{ourANNALS} Abdulla Rakhimov, Shuhrat Mardonov and  E.Ya. Sherman, Ann. Phys. {\bf 326}, 2499 (2011);\\
 Abdulla Rakhimov, E. Ya. Sherman and Chul Koo Kim,
Phys. Rev. B {\bf 81}, 020407(R) (2010).
\bi{Stoofmakola}D. van Osten, O. van der Straten and H.T.C. Stoof
Phys. Rev. A {\bf 63}, 053601 (2001).
\bi{ourYUK} V.I. Yukalov, A. Rakhimov, S. Mardonov, Laser Phys. {\bf 21}, 264 (2011).
\bi{KLVPT}  	H. Kleinert, S. Schmidt, and A. Pelster,
 Annalen der Physik (Leipzig) {\bf 14}, 214-230 (2005)
\bi{GFCM}
For field theories on a lattice see
 H. Kleinert,
     {\em Gauge Fields in Condensed Matter\/},
     Vol.~I \,\,  Superflow and Vortex Lines,
     World Scientific, Singapore 1989, pp. 1--756.
\bi{jackiw} R. Jackiw, Phys. Rev. D {\bf 9}, 1686 (1974).
\bi{oursinggap}	A. M.  Rakhimov, Jae Hyung Yee,  Intern. Journ. Mod. Phys. A {\bf 19}, 1589 (2004).
\bi{HK}
H. Kleinert, {\it
Converting Divergent Weak-Coupling into Exponentially Fast
Convergent}\\{\it Strong-Coupling Expansions},
EJTP {\bf8}, 25 (2011)\\
{\tt http://www.ejtp.com/articles/ejtpv8i25p15.pdf}.
\bi{danshita}I. Danshita and P. Naidon , Phys. Rev. A {\bf 79}, 043601  (2009).
\bi{andersen} J. O. Andersen, Rev. Mod. Phys.
{\bf 76}, 599 (2004).
 \bi{KS}
 {H. Kleinert} and
V. Schulte-Frohlinde,
    {\em Critical Phenomena in $\phi ^4$-Theory\/},
    World Scientific, Singapore, 2001.
\bi{kapustabellac} J. I. Kapusta {\it Finite Temperature Field Theory}, (Cambridge University Press, 1989);\\
 Michel Le Bellac  {\it Thermal Field Theory}, (Cambridge University Press, 1996).
\bi{haugset} T. Haugset, H. Haugerud and F. Ravndal,
	Ann. Phys. {\bf 27}, 266 (1998).
\bi{braaten} E. Braaten and A. Nieto, Euro. Phys. J. B {\bf 11}, 143 (1999).
\bi{yukannals} V. I. Yukalov Ann. Phys. {\bf 323}, 461 (2008)
\bi{Sheshadri} K. Sheshadri, H. R. Krishnamurthy, R. Pandit and T. V. Ramakrishnan,
Europhys. Lett. {\bf 22}, 257, (1993).
\bi{stev} P. M. Stevenson Phys. Rev. D
 {\bf 32}, 1389 (1985);\\
 Chul Koo Kim, A. Rakhimov and Jae Hyung Yee    Eur. Phys. Journ.  B {\bf 39}, 301 (2004);\\
 A. Rakhimov, Chul Koo Kim, Sang-Hoon Kim and Jae Hyung Yee, Phys. Rev. A {\bf 77}, 033626 (2008).
\bi{PI}
H. Kleinert,
{\it Path {I}ntegrals in {Q}uantum {M}echanics, {S}tatistics, {P}olymer {P}hysics, and {F}inancial {M}arkets}, 5th ed., World Scientific (2009).
   \eb
   \edc

\appendix B
{ Energy of the ground state and densities in operator formalism.}

Making Bogoliubov shift: $\hat{a}_{\sbf{i}}\to a_{\sbf{i}}+\sqrt{\nu  n_0}$ in Bose-Hubbard grand Hamiltonian we separate it as follows: \footnote{One may proceed this calculations in ordinary model, setting $\mu=\mu=\mu$}
\bea
 H&=&H_{0}+H_{1}+H_{2}+H_{3}+H_{4}\nonumber\\
  H_{0}&=&-J n_0z_{0}+\frac{U}{2}N_{s}\nu^{2} n_0^{2}-\mu N\nonumber\\
   H_{2}&=&-J\sum_{<{\sbf i},{\sbf j}>}a_{\sbf{i}}^{\ast}a_{\sbf{i}}+\frac{U}{2}\nu n_0\sum_{\sbf{i}}(\hat{S}_{1}+\hat{S}_{1}^{\ast})+[2U\nu  n_0-\mu]\hat{N}_{1}\nonumber\\
     H_{3}&=&U\sqrt{\nu  n_0}\sum_{\sbf j}[a_{\sbf{i}}^{\ast}a_{\sbf{i}}^{\ast}a_{\sbf{i}}+a_{\sbf{i}}^{\ast}a_{\sbf{i}}a_{\sbf{i}}]\nonumber\\
      H_{4}&=&\frac{U}{2}\sum_{\sbf{i}}a_{\sbf{i}}^{\ast}a_{\sbf{i}}a_{\sbf{i}}a_{\sbf{i}}\label{b1}
       \eea
   where $\hat{S}_{1}=\sum_{\sbf{i}}a_{\sbf{i}}a_{\sbf{i}}$ and $\hat{N}_{1}=\sum_{\sbf{i}}a_{\sbf{i}}^{\ast}a_{1}$ -- operators, we omitted $H_{1}$ term.

   Passing to the momentum space
   \begin{eqnarray}
    a_{1}&=&\frac{1}{\sqrt{N_{s}}}\sum_{{\sbf k}}\hat{a}_{{\sbf k}} \displaystyle{e^{i \bfk {\bf a}}},
      \qquad\sum_{{\sbf k}}1=N_{s},\nonumber\\
       &&[a_{{\sbf k}},p_{p}^{\ast}]=\delta(\bfk -{\bf p})
        \end{eqnarray}
where ${\bf a}=\{a_{\alpha}\}$ is a lattice vector and the quasimoment $\bfk$ pertains to the Brillouin zone, $-\pi/a_{\alpha}\leq k_{\alpha} \leq \pi/a_{\alpha}$ and using formulas () one may rewrite $H_{2}$ and $H_{4}$ as
\bea
 H_{2}&=&\sum_{{\sbf k}}\left[\varepsilon(k)+2U\nu  n_0-\mu-Jz_{0}\right]
  \hat{a}_{{\sbf k}}^{\ast}a_{{\sbf k}}+\frac{U\nu  n_0}{2}\sum_{{\sbf k}}[\hat{a}_{{\sbf k}}\hat{a}_{{\sbf k}}^{\ast}+
   a_{{\sbf k}}^{\ast}a_{{\sbf k}}^{\ast}]\nonumber\\
    H_{4}&=&\frac{U}{2N_{s}}\sum_{\bfk,{\bf p},{\bf q}}\hat{a}_{{\sbf k}}^{\ast}
     \hat{a}_{p}^{\ast}\hat{a}_{k+q}\hat{a}_{p-q}\label{b3}
      \eea
      with $\varepsilon(k)=2J\sum_{\alpha}[1-\cos k_{\alpha}a]>0$

In the Wannier representation the normal fraction reduces to
\bea
 n_{1}&\equiv&\frac{1}{N}\sum_{\sbf{i}}\langle a_{\sbf{i}}^{\ast}a_{\sbf{i}}\rangle=
  \frac{1}{N}\sum_{{\sbf k}}n_{{\sbf k}},\nonumber\\
   n_{{\sbf k}}&=&\langle a_{{\sbf k}}^{\ast}a_{{\sbf k}}\rangle\label{b4}
    \eea
 while the anomalous average becomes
\bea
 \sigma&\equiv&\frac{1}{N}\sum_{\sbf{i}}\langle a_{\sbf{i}}a_{\sbf{i}}\rangle=
  \frac{1}{N}\sum_{{\sbf k}}\sigma_{{\sbf k}},\nonumber\\
   \sigma_{{\sbf k}}&=&\langle a_{{\sbf k}}a_{{\sbf k}}\rangle \label{b5}
    \eea

 To diagonalize the Hamiltonian  we introduce phonon operators  $b_{{\sbf k}}, b_{{\sbf k}}^{\ast}$ through Bogoliubov transformation
 \bea
  \left.\begin{array}{ll}
   a_{{\sbf k}}=u_{{\sbf k}}b_{{\sbf k}}+v_{{\sbf k}}b_{{\sbf k}}^{\ast}\\
   a_{{\sbf k}}^{\ast}=u_{{\sbf k}}b_{{\sbf k}}^{\ast}+v_{{\sbf k}}b_{{\sbf k}}
    \end{array}\right\} \label{b6}
     \eea
where real variational function $u_{{\sbf k}}$ will be chosen so that it minimize the ground state energy:
\bea
 \cE_{{\rm gr}}&=&<H_{0}>+<H_{2}>+<H_{4}>,\nonumber\\
  \frac{\delta E}{\delta u_{{\sbf k}}}&=&0\label{b7}
   \eea
Now we must clarify "what do we mean by the ground state". In terms of Bogoliubov operators at zero temperature the ground state of the system  $| 0>$ is vacuum state for phonons, i.e.:
\bea
 b_{{\sbf k}} | 0>=0,\quad <0 |b_{{\sbf k}}^{\ast}=0\label{b8}
  \eea
where operators $b_{{\sbf k}}$  satisfy Bose commutation relations
\bea
 \left[b_{{\sbf k}}b_{{\sbf k}}^{\ast}\right]=\delta(\bfk-{\bf p})\nonumber
  \eea
Therefore we have set
\bea
 <b_{{\sbf k}}b_{p}>&=&<b_{{\sbf k}}^{\ast}b_{p}^{\ast}>=<b_{{\sbf k}}^{\ast}b_{p}>=0\nonumber\\
  <b_{{\sbf k}}b_{p}^{\ast}>&=&\delta(\bfk-{\bf p})\label{b9}
   \eea
 Inserting (\ref{b6}) into (\ref{b3})    and taking the average one may calculate the ground state energy of the Bose condensed system.

 To make easier such calculations we try to formulate some rules.

 First, lets consider $n_{1}$ and $\sigma$
 \bea
  n_{1}&=&\frac{1}{N}\sum_{{\sbf k}} n_{{\sbf k}}=\frac{1}{N}\sum<a_{{\sbf k}}^{\ast}a_{{\sbf k}}>=
   \frac{1}{N}\sum_{{\sbf k}}\Bigg\{u_{{\sbf k}}^{2}<b_{{\sbf k}}^{\ast}b_{{\sbf k}}>\nonumber\\
    &&{}+u_{{\sbf k}}v_{{\sbf k}}\left[<b_{{\sbf k}}^{\ast}b_{{\sbf k}}^{\ast}>+<b_{{\sbf k}}b_{{\sbf k}}>\right]+
     v_{{\sbf k}}^{2}<b_{{\sbf k}}b_{{\sbf k}}^{\ast}>\Bigg\}
      =\frac{1}{N}\sum_{{\sbf k}}v_{{\sbf k}}^{2}\label{b10}\\
   \sigma&=&\frac{1}{N}\sum \sigma_{{\sbf k}}=\frac{1}{N}\sum_{{\sbf k}}<a_{{\sbf k}}a_{{\sbf k}}>
    =\frac{1}{N}\sum_{{\sbf k}}\Bigg\{u_{{\sbf k}}^{2}<b_{{\sbf k}}^{\ast}b_{{\sbf k}}>+v_{{\sbf k}}^{2}
     <b_{{\sbf k}}^{\ast}b_{{\sbf k}}^{\ast}>\nonumber\\
      &&{}+u_{{\sbf k}}v_{{\sbf k}}\left[<b_{{\sbf k}}b_{{\sbf k}}^{\ast}>+<b_{{\sbf k}}^{\ast}b_{{\sbf k}}>\right]\Bigg\}
       =\frac{1}{N}\sum_{{\sbf k}}u_{{\sbf k}}v_{{\sbf k}}\label{b11}
        \eea
where we used (\label{b9})

>From (\ref{b10}) and (\ref{b11}) one may make sure that the pair contractions of operator $a_{{\sbf k}}$ are given as:
\bea
 \overline{a_{{\sbf k}}^{\ast}a_{p}^{\ast}}&=&\overline{a_{{\sbf k}}a_{p}}=
  u_{{\sbf k}}v_{p}\delta({\bf p}+\bfk)\nonumber\\
   \overline{a_{{\sbf k}}^{\ast}a_{p}}&=&v_{{\sbf k}}^{2}\delta(\bfk-{\bf p})\nonumber\\
    \overline{a_{p}a_{{\sbf k}}^{\ast}}&=&1+<a_{{\sbf k}}^{\ast}a_{{\sbf k}}>=
     1+v_{{\sbf k}}^{2}\delta(\bfk-{\bf p})=\delta(\bfk-{\bf p})[u_{{\sbf k}}^{2}]\label{b12}
      \eea

Now, having in hand these contractions we are able to calculate to calculate the average $\hat{A}(a_{1},a_{2},\ldots a_{1}^{\ast},a_{2}^{\ast}\ldots)$ using Wick theorem. For further reference we got from (\ref{b11}) and (\ref{b12}) following useful formulas:
\bea
 \left\{ \begin{array}{ll}
  \sum_{{\sbf k}}v_{{\sbf k}}^{2}=Nn_{1}\\
   \sum_{{\sbf k}}u_{{\sbf k}}v_{{\sbf k}}=N\sigma
  \end{array}\right.
   \eea
where $u_{{\sbf k}}, v_{{\sbf k}}$ will be  detailed further.

First we calculate $\cE^{(2)}$
\bea
 \cE^{(2)}&=&<H_{2}>=\sum_{{\sbf k}}\left[\varepsilon(k)+2U\nu  n_0-\mu-Jz_{0}
  \right]\overline{a_{{\sbf k}}^{\ast}a_{{\sbf k}}}\nonumber\\
   &&{}+ \frac{U\nu  n_0}{2}\sum_{{\sbf k}}\left[\overline{a_{{\sbf k}}a_{{\sbf k}}}
    +\overline{a_{{\sbf k}}a_{{\sbf k}}}\right]\nonumber\\
     &=&\sum_{{\sbf k}}\left[\varepsilon(k)+2U\nu  n_0-\mu-Jz_{0}\right]
      v_{{\sbf k}}^{2}\nonumber\\
       &&{}+U\nu  n_0\sum_{{\sbf k}}u_{{\sbf k}}v_{{\sbf k}}
        \eea
\edc

\begin{figure}[h]
\begin{minipage}[h]{0.49\linewidth}
\center{\includegraphics[width=1.0\linewidth]{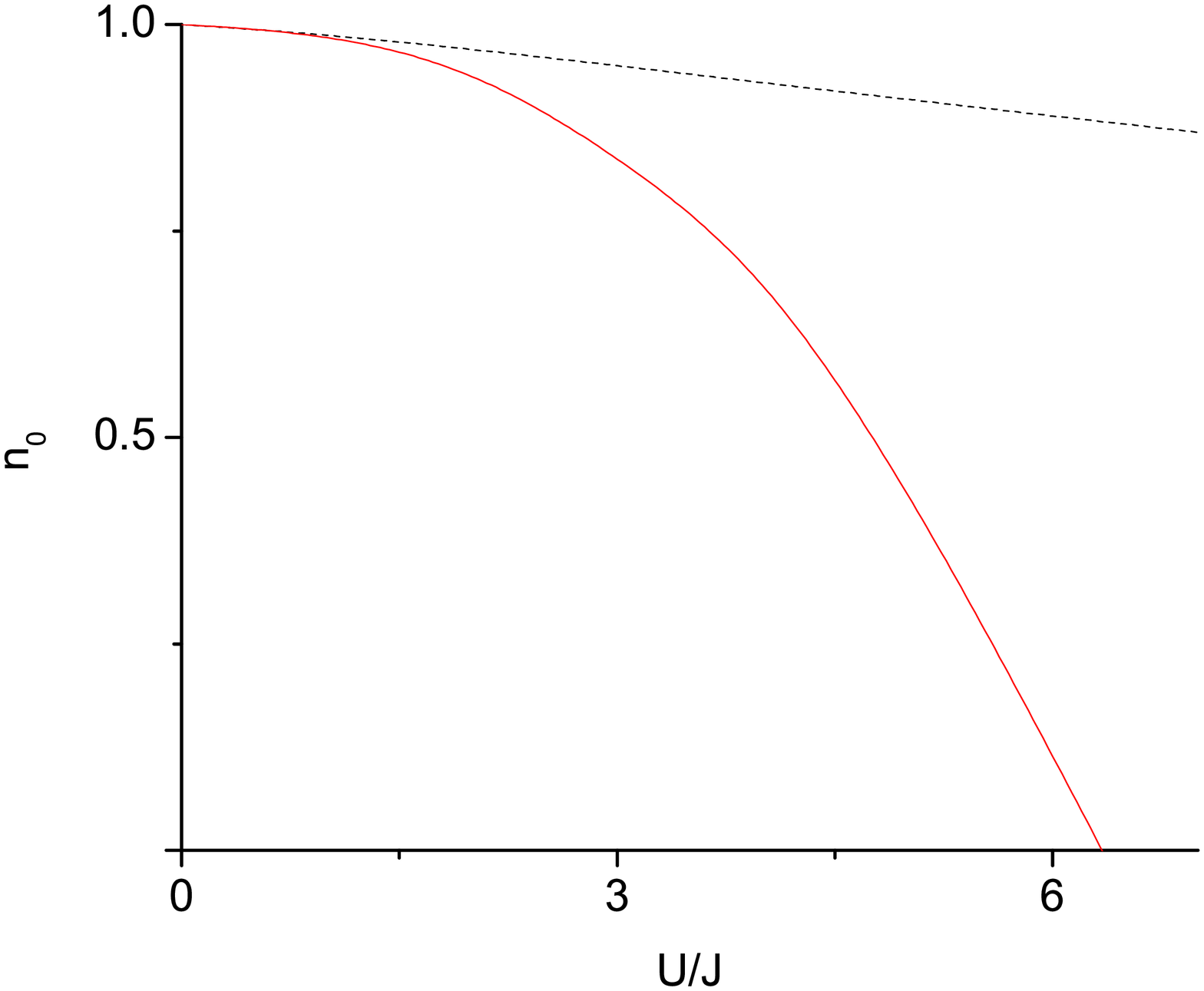} \\ a)}
\end{minipage}
\hfill
\begin{minipage}[h]{0.49\linewidth}
\center{\includegraphics[width=1.0\linewidth]{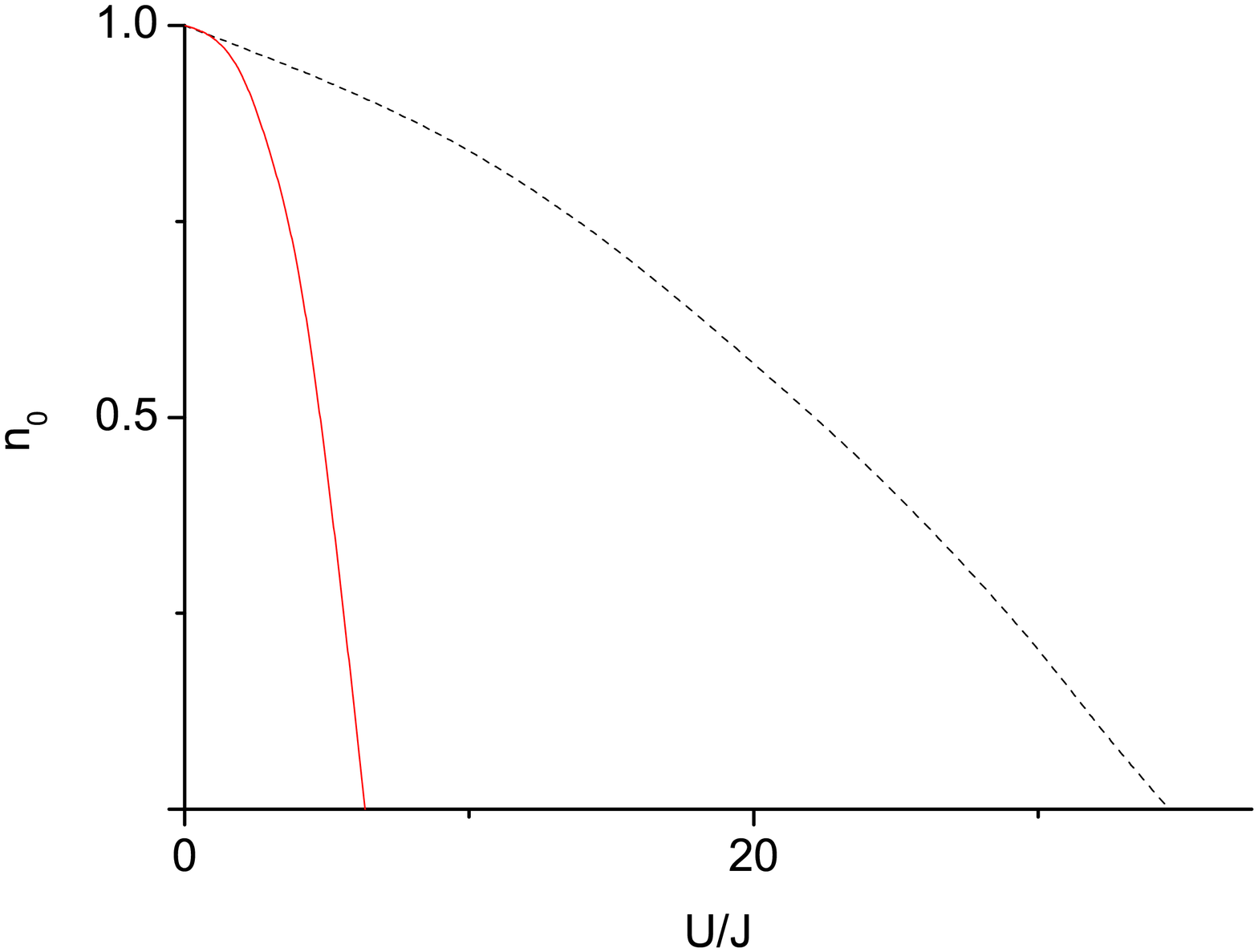} \\ b)}
\end{minipage}
\caption{The superfluid fraction $n_0$ as a function of  $U/J$ for $\nu=1$, $D=3$.
a)In one  and two loop approximations  (dashed line and solid lines, respectively); b)
Here the dashed  line was  obtained in Gutzwiller approach
 while the solid line in the  present one.}
\label{fig:2}
\end{figure}